\documentclass{aa}

\usepackage{hyperref}
\usepackage{pdflscape}
\usepackage{xcolor}
\usepackage{amsmath,bm}

\usepackage{graphicx}
\usepackage{txfonts}

\begin{document}

   \title{The young massive SMC cluster NGC~330 seen by MUSE.
          \thanks{Based on observations collected at the ESO Paranal
          observatory under ESO program 60.A-9183(A).}}
   \subtitle{I. Observations and stellar content}

   \author{J. Bodensteiner\inst{1},
   H.~Sana\inst{1},
   L.~Mahy\inst{1},
   L.~R.~Patrick\inst{2,3},
   A.~de Koter\inst{1,4},
   S.~E.~de~Mink\inst{5, 4},
   C.~J.~Evans\inst{6},
   Y.~Götberg\inst{7},
   N.~Langer\inst{8},
   D.~J.~Lennon\inst{2},
   F.~R.~N.~Schneider\inst{9},
   F.~Tramper\inst{10}
   }

   \institute{Instituut voor Sterrenkunde, KU Leuven, Celestijnenlaan 200D, 3001 Leuven, Belgium\\
              \email{julia.bodensteiner@kuleuven.be}
    \and
    Instituto de Astrof\'isica de Canarias, E-38205 La Laguna, Tenerife, Spain
    \and
    Universidad de La Laguna, Dpto. Astrof\'isica, E-38206 La Laguna, Tenerife, Spain
    \and
    Astronomical Institute Anton Pannekoek, Amsterdam University, Science Park 904, 1098 XH, Amsterdam, The Netherlands
    \and
    Center for Astrophysics, Harvard $\&$ Smithsonian, 60 Garden Street, Cambridge, MA 02138, USA
    \and
    UK Astronomy Technology Centre, Royal Observatory, Blackford Hill, Edinburgh, EH9 3HJ, UK
    \and
    The observatories of the Carnegie institution for science, 813 Santa Barbara St., Pasadena, CA 91101, USA
    \and
    Argelander-Institut für Astronomie, Universität Bonn, Auf dem Hügel 71, 53121 Bonn, Germany
    \and
    Heidelberger Institut für Theoretische Studien, Schloss-Wolfsbrunnenweg 35, 69118 Heidelberg, Germany
    \and
    Institute for Astronomy, Astrophysics, Space Applications $\&$ Remote Sensing, National Observatory of Athens, P. Penteli, 15236 Athens, Greece}

   \date{Received 19 September 2019; accepted 6 November 2019}

  \abstract
   {A majority of massive stars are part of binary systems, a large fraction of which will inevitably interact during their lives. Binary-interaction products (BiPs), i.e. stars affected by such interaction, are expected to be commonly present in stellar populations. BiPs are thus a crucial ingredient in the understanding of stellar evolution.}
   {We aim to identify and characterize a statistically significant sample of BiPs by studying clusters of $10-40$\,Myr, an age at which binary population models predict the abundance of BiPs to be highest. One example of such a cluster is NGC\,330 in the Small Magellanic Cloud.}
   {Using MUSE WFM-AO observations of NGC 330, we resolve the dense cluster core for the first time and are able to extract spectra of its entire massive star population. We develop an automated spectral classification scheme based on the equivalent widths of spectral lines in the red part of the spectrum.}
   {We characterize the massive star content of the core of NGC~330 which contains more than 200 B stars, two O stars, 6 A-type supergiants and 11 red supergiants. We find a lower limit on the Be star fraction of $32 \pm 3\%$ in the whole sample. It increases to at least $46 \pm 10\%$ when only considering stars brighter than $V=17\,\mathrm{mag}$.
   We estimate an age of the cluster core between 35 and 40 Myr and a total cluster mass of $88^{+17}_{-18} \times 10^3 M_{\odot}$.}
   {We find that the population in the cluster core is different than the population in the outskirts: while the stellar content in the core appears to be older than the stars in the outskirts, the Be star fraction and the observed binary fraction are significantly higher. Furthermore, we detect several BiP candidates that will be subject of future studies.}

   \authorrunning{Bodensteiner et al.}

   \maketitle
%

%
\section{Introduction}\label{Sec:intro}
Massive stars, i.e. stars with birth masses larger than $8M_{\odot}$, have played a major role throughout the history of the Universe \citep{Jamet2004, Bresolin2008, Aoki2014}. Despite their low numbers and short lifetimes, they have a large impact on their environment. They heat, shape and enrich their surroundings with their high temperatures and luminosities, strong stellar winds, and final explosions \citep{Bromm2009, DeRossi2010}. Massive stars can evolve following a diversity of evolutionary channels. Which channel a star follows depends most significantly on its initial mass, metallicity, rotation, and multiplicity. \newline
Recent observations have shown that a vast majority of massive stars do not evolve as single stars \citep[see e.g.,][]{Sana2012, Kobulnicky2014, Dunstall2015}. In contrast, they reside in binary systems and will, at some point in their lives, interact with a nearby companion \citep{Sana2012}. Besides stable mass transfer in terms of Roche-lobe overflow, more dramatic interactions such as common-envelope evolution and stellar mergers are possible \citep{Taam2000, deMink2014, Kochanek2019}. In any of these cases, the star's mass and angular momentum change significantly which will drastically alter the evolutionary path of the two stars \citep{Podsiadlowski1992, Vanbeveren1994}. Understanding the detailed evolution of massive stars, taking into account the effect of multiplicity, remains one of the biggest challenges in modern astrophysics \citep{Langer2012}.\newline
One possible line of investigation in order to explore the effect of multiplicity on the evolution of massive stars is to study stars after they have interacted, i.e. binary-interaction products (BiPs). Indeed, the outcome of the interaction process, which defines the physical properties of such BiPs, remains ill-constrained. Characterizing the physical properties of BiPs is, however, crucial to assess their current impact on their surroundings and predict their future evolution. This in turn defines the nature and quantity of key end-of-life products, such as supernova types and the strength of supernova kicks, the nature and multiplicity properties of compact objects, and ultimately, those of gravitational wave events. \newline
BiPs can be identified through various stellar properties, unfortunately none of which ensure an unambiguous identification. Due to the transfer of angular momentum, mass gainers or merger products may be expected to be rapidly rotating \citep[][Mahy et al. 2019, subm.]{deMink2013}. The transfer of mass may lead to abnormal chemical abundances on the stellar surface \citep{deMink2007, Langer2012}. The supernova explosion of the more massive star can disrupt the system, turning the secondary into a so-called runaway star, a star moving away from its birth place with a high space velocity \citep{Gies1986, Blaauw1993, Hoogerwerf2001}. If the system remains bound, an excess in UV or hard X-ray flux may indicate the presence of a compact companion like in the case of subdwarf companions \citep[see e.g. ][]{Gies1998, Wang2017, Wang2018} and X-ray binaries \citep{Verbunt1993, Reig2011}. In star clusters, mass-gainers and mergers are expected to appear younger than their sibling stars, i.e. they are bluer and hotter than the main-sequence turn-off and are massive analogs to the Blue Stragglers found in globular clusters \citep[see e.g.,][]{Sandage1953, Shara1997, Ferraro1997, Schneider2015}. Recently, their cooler equivalent was proposed by \citet{Britavskiy2019} and \citet{Beasor2019}: so-called Red Stragglers, i.e. red supergiants (RSGs) that are more massive and luminous than expected from single-star evolution.\newline
Next to the study of the well-known post-mass transfer Algol systems, searches for BiPs were conducted in several manners in the past: numerous photometric studies found Blue Stragglers in all types of star clusters \citep[see e.g.,][]{Ferraro1997, Ahumada2007, Gosnell2014}. \cite{Schneider2014} reported indications for BiP signatures in the mass function of young clusters while \cite{Ramirez2013, Ramirez-Agudelo2015} found a significant proportion of fast rotators in the presumably single O-type stars in 30 Dor. The relative fraction of these were compatible with population synthesis predictions of BiPs \citep{deMink2013, deMink2014}. The search for BiPs was extended to magnetic stars through the Binary and Magnetism survey \citep[BinaMIcS,][]{Alecian2015, Schneider2016}. \cite{Bodensteiner2018} proposed to use infrared nebulae around massive stars possibly formed in previous binary interactions as general tracers for BiPs. \newline
Population synthesis simulations show that the relative contributions of BiPs to the lower mass content of the cluster vary as a function of age, and peak between $8 - 40$\,Myrs, assuming a single starburst \citep{Schneider2015}. While much effort has been invested in spectroscopically characterizing the massive star content of young massive star clusters \citep[see e.g., ][and references therein]{Evans2011, Sana2011, Clark2014, Ramirez-Tannus2018}, these slightly older populations have been relatively overlooked. \newline
Here we aim to investigate the massive star population of the young open star cluster NGC~330. This cluster is situated in the Small Magellanic Cloud (SMC) at a distance of $60\pm1\,\mathrm{kpc}$ \citep{Harries2003, Hilditch2005, Deb2010}. Age estimates of NGC~330 vary between 26 and 45 Myrs, thus putting it in the age range of interest for our purposes \citep[][Patrick et al., subm.]{Sirianni2002, Martayan2007a, Milone2018}.
The estimated photometric mass of NGC~330 ranges between $M_\mathrm{phot} \approx 3.6 \times 10^4\, M_{\odot}$ \citep{McLaughlin2005} and $M_\mathrm{phot} \approx 3.8 \times 10^4\, M_{\odot}$ \citep{Mackey2002} which is in good agreement with a recent estimate of the dynamical mass $M_\mathrm{dyn} = 15.8^{+7.6}_{-5.1} \times 10^4 \,M_{\odot}$ (Patrick et al., subm.). \newline
Several spectroscopic studies of the brightest stars in NGC~330 were conducted in the past reporting a high Be star fraction of about $50 \%$ or higher \citep{Feast1972, Grebel1992, Lennon1993, Grebel1996, Mazzali1996, Keller1998, Keller1999}. Based on these findings, \citet{Maeder1999} concluded that the Be star fraction in the SMC is higher than the Large Magellanic Cloud and in the Milky Way. A high Be star fraction in SMC clusters (excluding NGC\,330) was confirmed by \citet{Martayan2007a}. Around 150 OB stars situated in the outskirts of NGC~330 were studied spectroscopically \citep{Evans2006}.
Additionally, the cluster was recently investigated by \citet{Milone2018} using Hubble Space Telescope (HST) photometry. The dense cluster core containing more than 100 massive stars could not be resolved with existing spectroscopic instrumentation. It is, however, the region where, apart from runaways, many of the massive BiPs are expected due to mass segregation, be it primordial \citep{Sirianni2002} or as a result of cluster dynamics \citep{Portegies2010}.\newline
In this first paper of a series, we report on a spectroscopic study of the core of NGC~330 with the Multi-Unit Spectroscopic Explorer (MUSE) mounted at the VLT/UT4 \citep{Bacon2010}. With its capabilities enhanced with adaptive optics (AO), it is now possible to efficiently obtain spectroscopy of individual stars in the dense cluster core. This first paper focuses on the methodology for spectral extraction and on the stellar content of the cluster. Future papers will address multiplicity and physical properties of the stars in the cluster. In Sect.\,\ref{Sec:obs} we summarize the observations and the data reduction. Sect.\,\ref{Sec:specextract} describes the automatic routine we develop to extract spectra of stars brighter than $V\,=\,18.5\,\mathrm{mag}$. In Sect.\,\ref{Sec:specclass} we explain how to automatically determine spectral types based on the equivalent widths of certain spectral lines between 4600 $\AA$ and 9300$\AA$. We present the stellar content of the cluster core in Sect.\,\ref{Sec:results} and compare it to previous photometric and spectroscopic studies of the cluster. Sect.\,\ref{Sec:conclusion} gives a summary and conclusions.

\section{Observations and Data Reduction}\label{Sec:obs}
\subsection{Observations with MUSE-WFM-AO}\label{Sec:MUSE}
\begin{figure*} \centering
  \includegraphics[width=0.99\hsize]{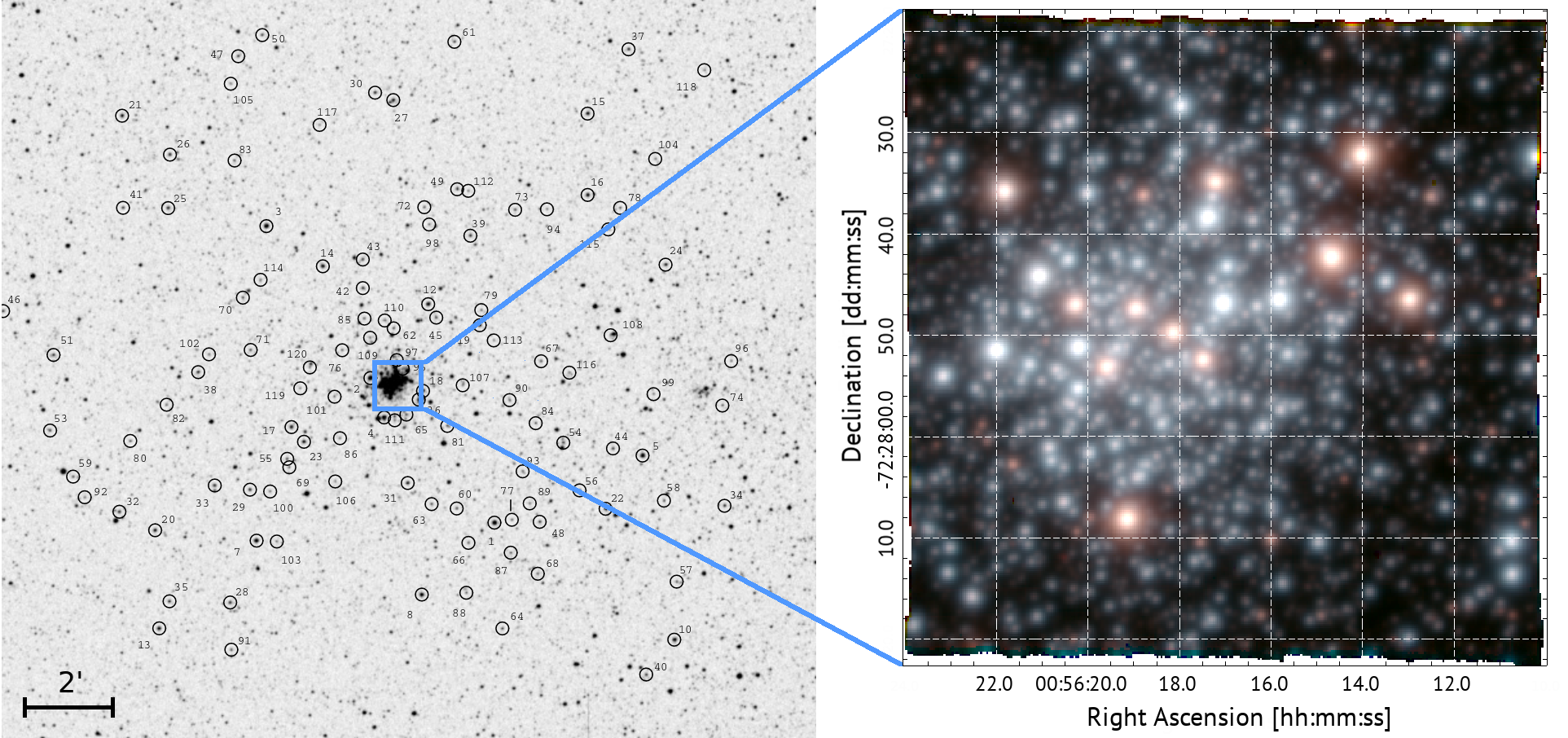}
  \caption{Left: Finding chart from \citet{Evans2006}, where stars that were studied are marked with black circles and the FoV of MUSE is indicated in blue. Right: True color image created by convolving the MUSE data with a $B$, $V$, and $I$ filter. At the distance of NGC~330, the size of the FoV of 1'$\times$1' corresponds to $\sim$ 20 $\times$ 20\,pc.}
  \label{fig:rgb}
\end{figure*}
NGC~330 was observed with MUSE on November $19^\mathrm{th}$, 2018, at Yepun, one of the four Unit Telescopes of the Very Large Telescope (VLT) in Paranal, Chile. MUSE is an integral field spectrograph comprising 24 individual spectrographs.\newline
In wide-field mode (WFM), the 24 individual spectrographs sample a Field of View (FoV) of 1'$\times$1' with a spatial sampling of 0.2'' in both directions.
The seeing during the observations varied between 0.4" and 0.7".
With the new AO, the MUSE-WFM observations are supported by ground-layer adaptive optics (GLAO) consisting of a deformable secondary mirror, four sodium laser guide stars and the Ground Atmospheric Layer Adaptive OptiCs for Spectroscopic Imaging (GALACSI) module. The observations were carried out in the extended wavelength mode covering a broad optical range (i.e. 4600$\AA$ -- 9300$\AA$). When supported by AO, the region between 5780 and 5990 $\AA$ is blocked by a notch filter in order to avoid light contamination by the sodium lasers of the AO system. The spectral resolving power ranges from 2000 at $4600\AA$ to 4000 at $9300\AA$.\newline
Five dither positions (DPs) of 540 seconds each were obtained with a relative spatial offset of about 0.7''. The instrument derotator was offset by 90$^{\circ}$ after each DP so that the light of each star was dispersed by multiple spectrographs. Directly after the observation of NGC~330 an offset sky observation for sky subtraction was taken (see Sect. \ref{Sec:sky}). A comparison between the finding chart of \citet{Evans2006} and a true color image showing the spatial coverage of the MUSE observations is shown in Fig.~\ref{fig:rgb}. \newline

\subsection{Data reduction}\label{Sec:reduc}
The MUSE data were reduced with the standard ESO MUSE pipeline v2.6\footnote{\textsc{https://www.eso.org/sci/software/pipelines/muse/}}. The calibrations for each individual IFU included bias and dark subtraction, flat fielding, wavelength and illumination correction. After re-combining the data from each IFU to the merged data cube, a telluric correction and a flux calibration was performed with the help of a standard star observation. Furthermore the sky was subtracted in different manners to obtain the best possible result (see Sect. \ref{Sec:sky}). The output of the data reduction is a so-called 3D data cube containing two spatial and one spectral direction of dimensions 321 $\times$ 320 $\times$ 3701. This can either be thought of as 3701 monochromatic images, or about $10^5$ individual spectra.\newline
In previous versions of the ESO MUSE pipeline (up to v2.4.1) large-scale (i.e. about $50 \AA$ wide) wiggles dominated the blue part of the spectrum when using the MUSE-WFM-AO setup. While this problem was fixed in the data extension itself in v2.6, it still occurs in the variance of the reduced MUSE data.

\subsection{Sky subtraction}\label{Sec:sky}
\begin{figure*} \centering
\includegraphics[width=1.\hsize]{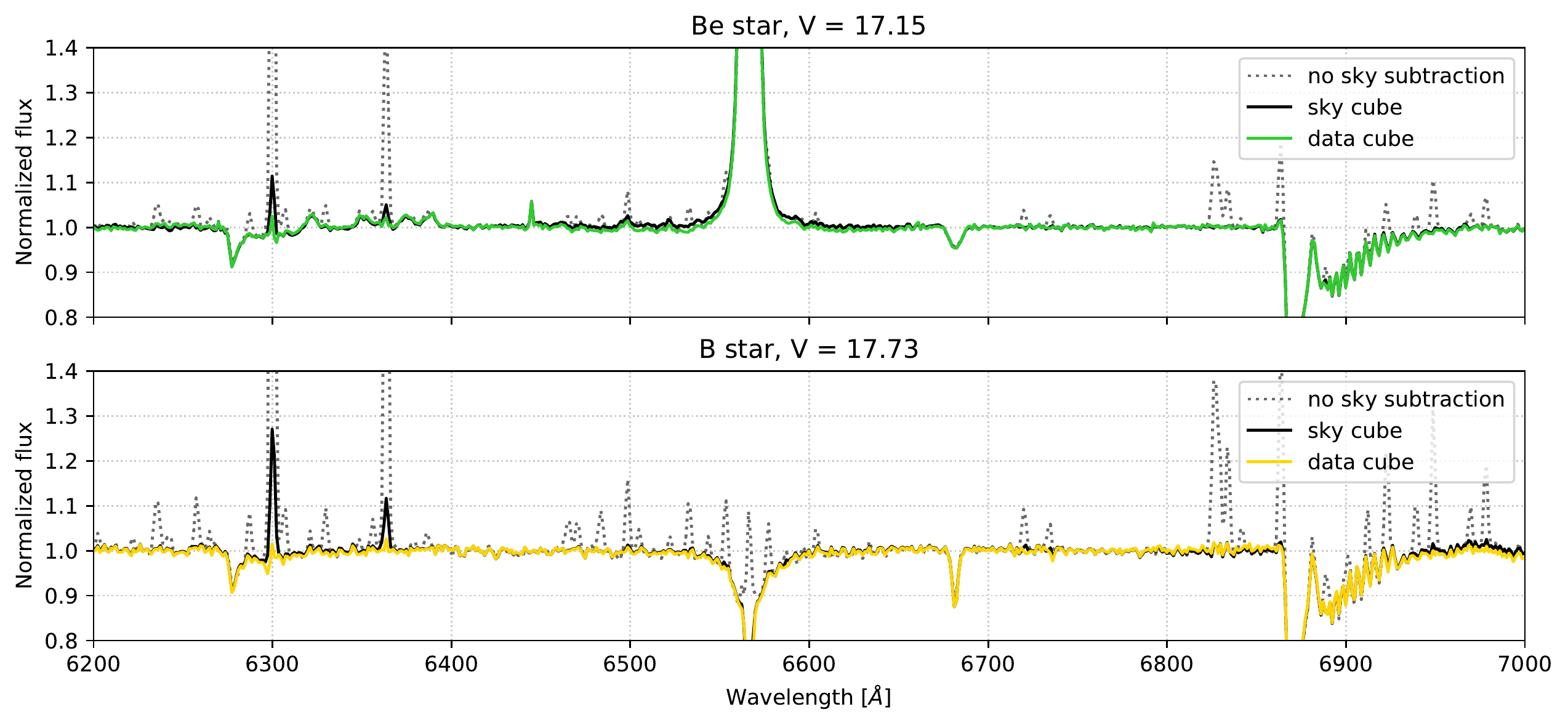}
\caption{Comparison of the sky subtraction methods for two example stars (marked in Fig. \ref{fig:hst_final} with the corresponding colors). The top panel gives the spectra of an example bright star (\#51, a Be star with $V=17.15\,\mathrm{mag}$) and the bottom panel gives those of a fainter star (\#33, a B star with $V=17.51\,\mathrm{mag}$). For both stars the original spectrum without any sky subtraction (no sky subtraction), the spectrum with the sky subtraction deduced from the dedicated sky cube (sky cube), and the spectrum with the sky subtraction determined in the science data cube (data cube) are shown. The spectra are not corrected for telluric lines (see e.g. the region between 6850 and 7000\,\AA).}
\label{fig:sky_subtraction}
\end{figure*}

There are different methods to subtract the sky in the ESO MUSE pipeline. As described in Sect.\,\ref{Sec:MUSE}, an offset sky exposure was taken at the end of the observational sequence. It is supposed to be centered on an effectively empty sky region but close to the target region. This facilitates the detection of a dark patch of sky that is not contaminated by stars in order to determine an average sky spectrum. While this procedure works well for isolated objects like galaxies, in the case of NGC~330, which is situated in a dense region of the SMC, it is difficult to find a nearby empty region. In this case, the sky observation is only 3' away of the science observation and is equally crowded by SMC stars so that the benefits of the sky observation are limited. \newline
Extensive testing showed that the remaining sky residuals are smaller when estimating the sky background from the science observations themselves. The MUSE pipeline offers the required framework in which a fraction of the darkest pixels in the science observation is used to estimate an average sky spectrum. This is then subtracted from the science data. The advantage of this method in comparison to using the sky cube is that the average sky spectrum is taken at exactly the same time and thus under the exact same weather and atmosphere conditions.\newline
A comparison between the two different sky subtraction methods for a bright star (top panel) and a faint star (bottom panel) is shown in Fig.\,\ref{fig:sky_subtraction}. The figure illustrates that (1) sky subtraction is an important task in the data reduction, and (2) that the residual sky emission lines are smaller using the data cube rather than using the sky cube (see e.g. at $\lambda \lambda$ 6300, 6365 $\AA$).

\section{Extraction of stellar spectra}\label{Sec:specextract}
\subsection{Targets input catalog}\label{Sec:inputlist}
\begin{figure} \centering
\includegraphics[width=0.99\hsize]{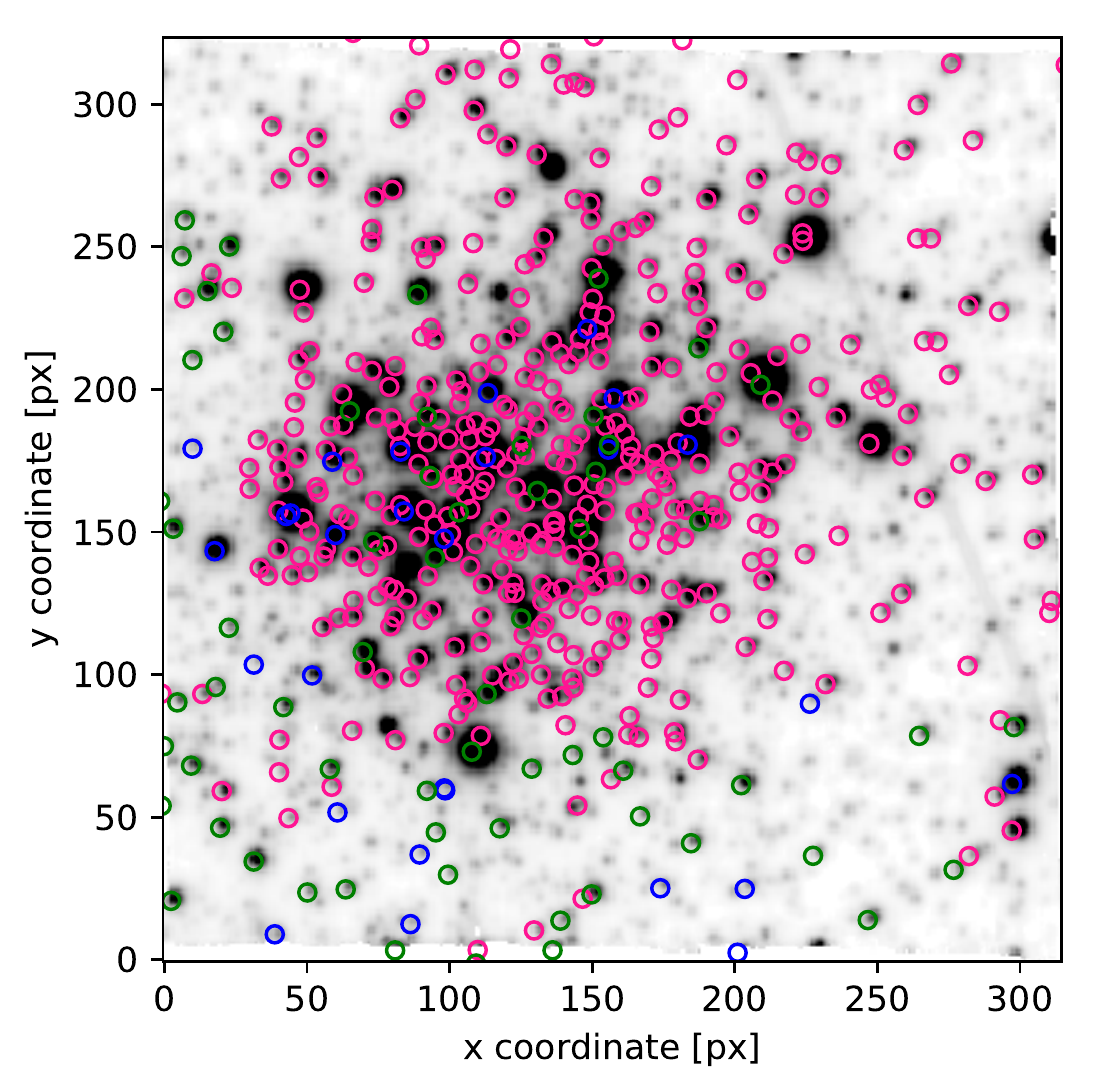}
\caption{HST source positions overlaid over an integrated-light image extracted from the MUSE cube. Stars with only F336W magnitudes are marked in pink while stars with only F555W magnitudes are marked in blue. Stars which appear in both lists and are brigther than 19.5 in F555W are marked in green. The latter ones are used to determine a correlation between the two magnitudes (see Fig. \ref{fig:hst_correlation}). The two HST source catalogs need to be aligned with respect to the MUSE coordinate system by shifting them by $\Delta x = (1.79\pm0.36)$ px and $\Delta y =(1.06\pm0.43)$ px, and rotating them by $\theta = (-0.05 \pm 0.10)^{\circ}$.}
\label{fig:hst_twoinlists}
\end{figure}

\begin{figure} \centering
\includegraphics[width=0.99\hsize]{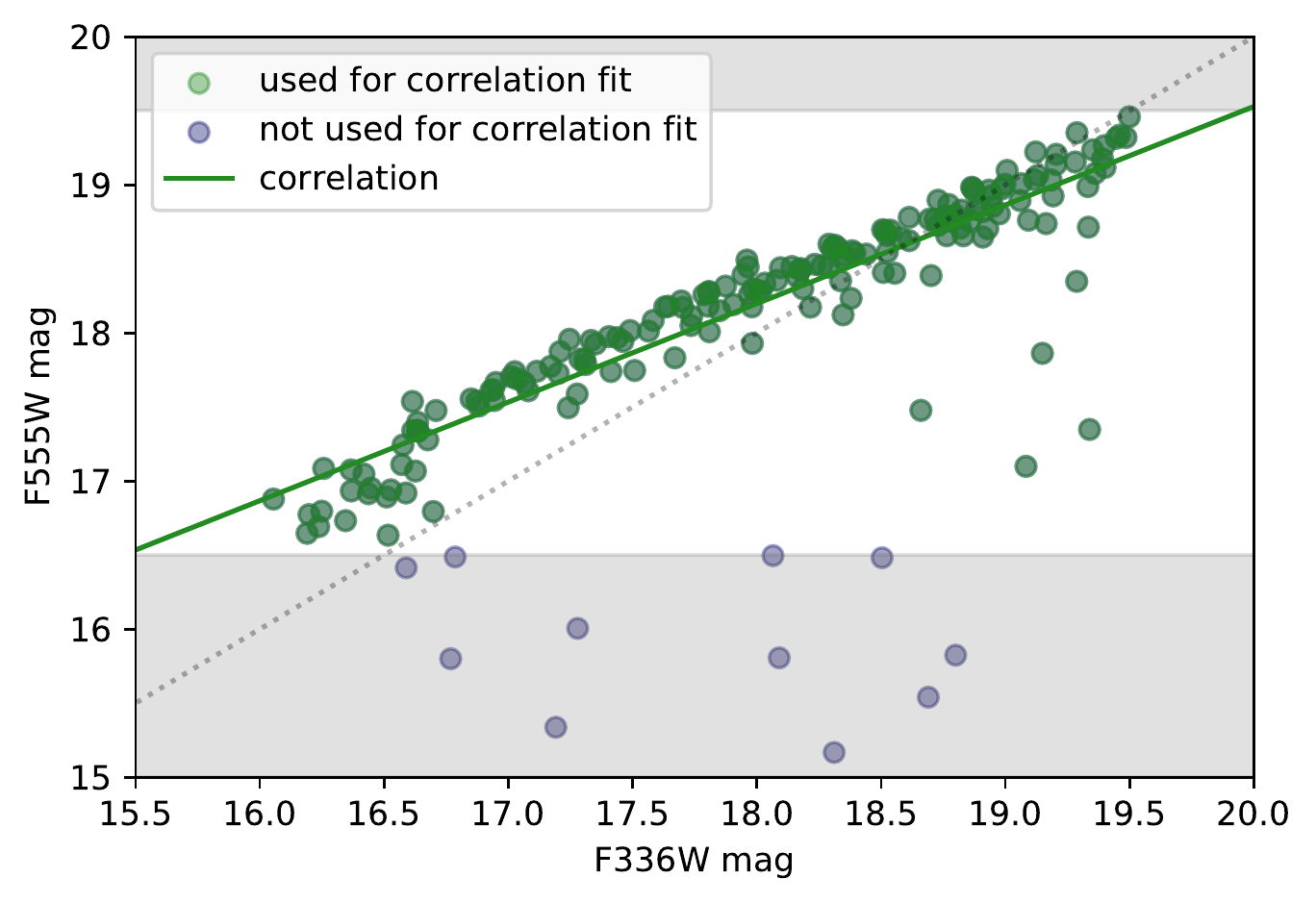}
\caption{Correlation between the HST F336W ($\sim$ $U$-band) and F555W ($\sim$\,$V$-band) magnitudes. In order to fit the linear correlation (green line) we exclude stars with F555W fainter than 19.5 (see Sec. \ref{Sec:inputlist}) and  brighter than 16.5 (gray regions). Stars that are brighter in F555W but not in the UV are RSGs and thus not included in the fit. The relation is used to convert F336W magnitudes to F555W magnitudes for all stars in the HST input catalogs. The grey dotted line indicates the one-to-one relation.}
\label{fig:hst_correlation}
\end{figure}

To extract spectra from the MUSE data, we require an input list of stars including stellar positions and brightnesses. We adopt a magnitude cut at $V\,=\,18.5\,\mathrm{mag}$ which approximately corresponds to $6M_{\odot}$ at SMC distance and extinction. This results in a signal-to-noise ratio (S/N) of around 80 which is high enough for our further analysis.\newline
Accurate stellar positions can be estimated in the MUSE data directly. They are, however, preferably determined from an instrument of higher spatial resolution, the Hubble Space Telescope (HST), providing an approximately ten times better spatial resolution. The core of NGC\,330 was observed with WFPC2 and the F555W filter (approximately corresponding to the $V$-band) in 1999, and with WFPC3 and the F336W filter (approximately corresponding to the $U$-band) in 2015. Unfortunately, the F555W filter image does not cover the full MUSE FoV (see Fig. \ref{fig:hst_twoinlists}).
The F336W data cover the whole FoV but the filter is outside the wavelength range covered by MUSE. For both images, an automatically created source list based on DAOphot is available on the HST archive\footnote{\textsc{https://archive.stsci.edu/hst/}}. The brightest stars (i.e. stars brighter than $V<15\,\mathrm{mag}$) are saturated in the HST images and thus missing from the input list.\newline
As can be seen in Fig. \ref{fig:hst_twoinlists}, the coordinate systems in both HST input catalogs are not aligned with the MUSE world coordinate system (WCS). We thus first align the two HST catalogs with each other, and in a subsequent step with the MUSE WCS. For this, we pick a handful of isolated stars in the MUSE image, determine their positions by fitting a point-spread function (PSF), and then compute the coordinate transformation between these known sources in MUSE and in the aligned HST catalogs. The coordinate transformation from the HST coordinate system (x', y') into the MUSE coordinate system (x, y) is given by:
\begin{flalign}
  x & = +x' \cdot \cos(\theta) + y' \cdot \sin(\theta) + \Delta x \\
  y & = -x' \cdot \sin(\theta) + y' \cdot \cos(\theta) + \Delta y
\end{flalign}
The best-fit values derived for the coordinate transformation are $\Delta x = (1.79\pm0.36)$ px, $\Delta y =(1.06\pm0.43)$ px, and $\theta =(-0.05 \pm 0.10) ^{\circ}$. This transformation is applied to all HST sources.\newline
To obtain a complete target list containing $V$ magnitudes of all stars we combine the F336W and the F555W list. We use stars appearing in both lists and brigther than 19.5 in F555W to compute a correlation between F555W magnitudes and F336W magnitudes. This allows us to approximately convert F336 magnitudes in F555W magnitudes for stars which are not covered by the F555W data (see Fig. \ref{fig:hst_twoinlists} and \ref{fig:hst_correlation}). This conversion is only accurate for early type main-sequence stars, for which the $U$-band filter covers the Rayleigh-Jeans tail of the spectrum. It does not hold for stars that are evolved off the main sequence (MS), e.g. for RSGs and blue supergiants (BSGs). As mentioned above, given the brightness of these stars, they are saturated in the HST images. For the same reason they have, on the other hand, been previously studied \citep[see e.g.][]{Arp1959, Robertson1974, Grebel1996}. We thus adopt their $V$-band magnitudes from \citet{Robertson1974} when available (see Tab. \ref{tab:RSGBSG}). \newline
The final target list is constructed by taking all sources that are within the MUSE field of view and brighter than $V=18.5$\,mag. Seven stars that are clearly brighter than $V=18.5$\,mag are missing in both HST input lists. This is probably due to problems in the automated source finding routines. These seven stars are added manually before constructing the final target list which contains 278 stars.
\begin{figure} \centering
\includegraphics[width=0.99\hsize]{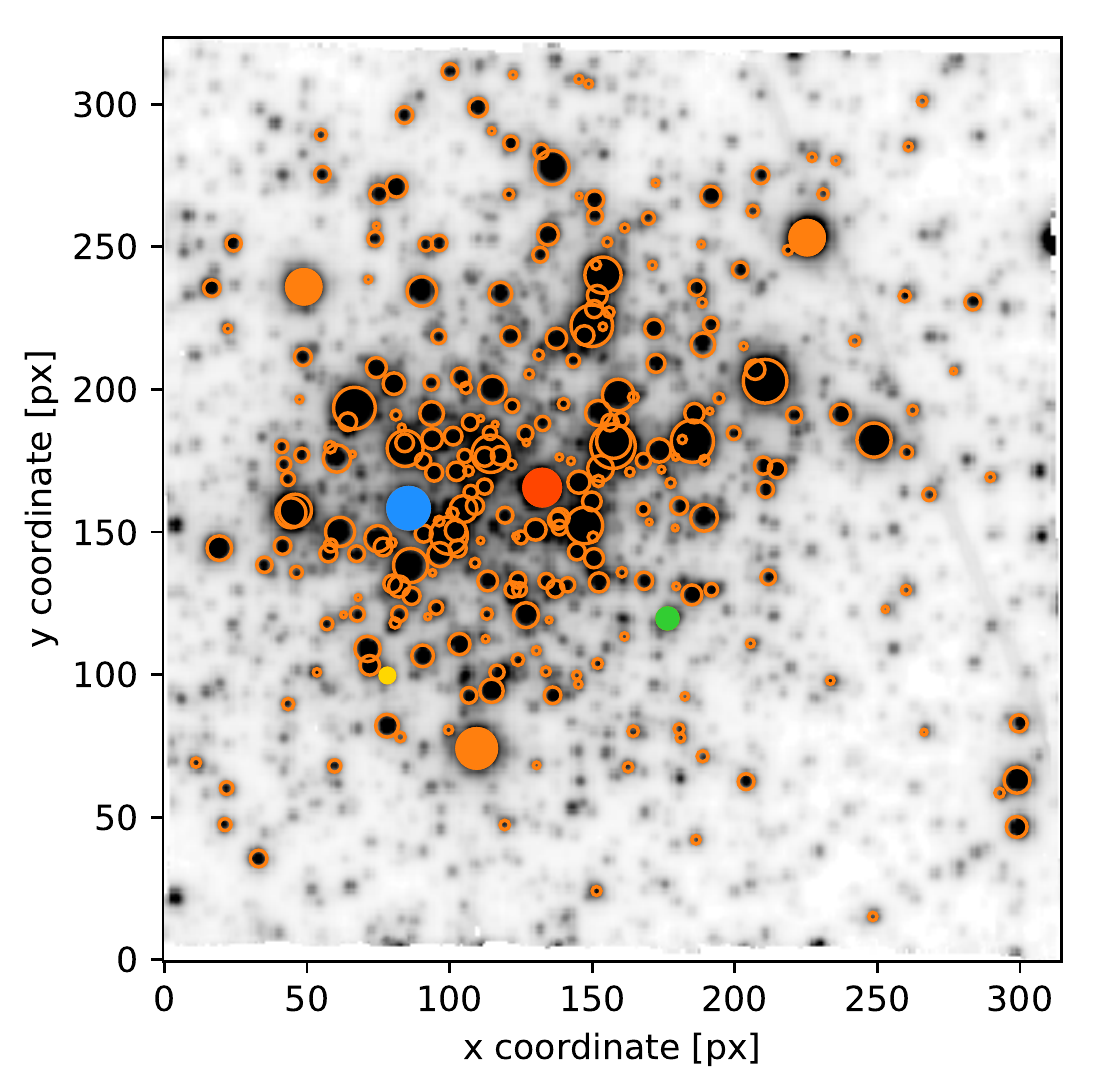}
\caption{Stars contained in the final target list (i.e. stars brigther than 18.5 in F555W) overplotted in orange over the MUSE white light image. The size of the circle corresponds to their inferred brightness in the $V$-band. The stars marked with orange filled circles are used to fit the PSF (see Sect. \ref{Sec:source_extraction}). The spectra of the stars marked with red, blue, green, and yellow filled circles are shown in Fig.\,\ref{fig:example_specs}.}
\label{fig:hst_final}
\end{figure}

\subsection{Source extraction}\label{Sec:source_extraction}
The high source density especially in the center of the image and the associated spatial overlap of sources makes extracting stellar spectra a complicated task. Given the format of the reduced MUSE data (i.e., a 3D data cube with two spatial and one spectral dimension), obtaining a spectrum of a source is equivalent to measuring its brightness in each wavelength image (referred to as a slice in the following). We avoid simple aperture photometry (i.e. adding up the flux in a given aperture around each star's position) in each slice because of the crowding of the field. \newline
We develop our own spectral extraction routine that employs PSF fitting. The routine is mainly based on the \textsc{python} package \textsc{photutils} \citep{Bradley2019}, a translation of DAOphot \citep{Stetson1987} into \textsc{python}, which provides a framework for PSF fitting in crowded fields. We use the input list described in Sect. \ref{Sec:inputlist} containing the 278 brightest stars. Subsequently, an effective PSF model \citep{Anderson2000} is derived from a selected set of isolated stars (see Fig. \ref{fig:hst_final}). Slice by slice (i.e. wavelength by wavelength), this PSF model is fitted to each source, while the positions taken from the input list are held fixed. In this way, we determine the total flux and flux error of a given star in each slice. All fluxes and error values for each star are then concatenated to a spectrum. \newline
 Crowding of the field is taken into account by simultaneously fitting the PSF of stars that are closer together than 12 pixels (corresponding to 2.4"). An example of the extraction of spectra for deblended sources is shown in Appendix \ref{Sec_appendix:specextract}. For seven stars the extraction of a spectrum was not possible due to severe crowding. In total we extract spectra for 271 stars brighter than $V=18.5\,\mathrm{mag}$.

\subsection{Normalization}\label{Sec:norm}
\begin{figure*} \centering
  \includegraphics[width=0.99\hsize]{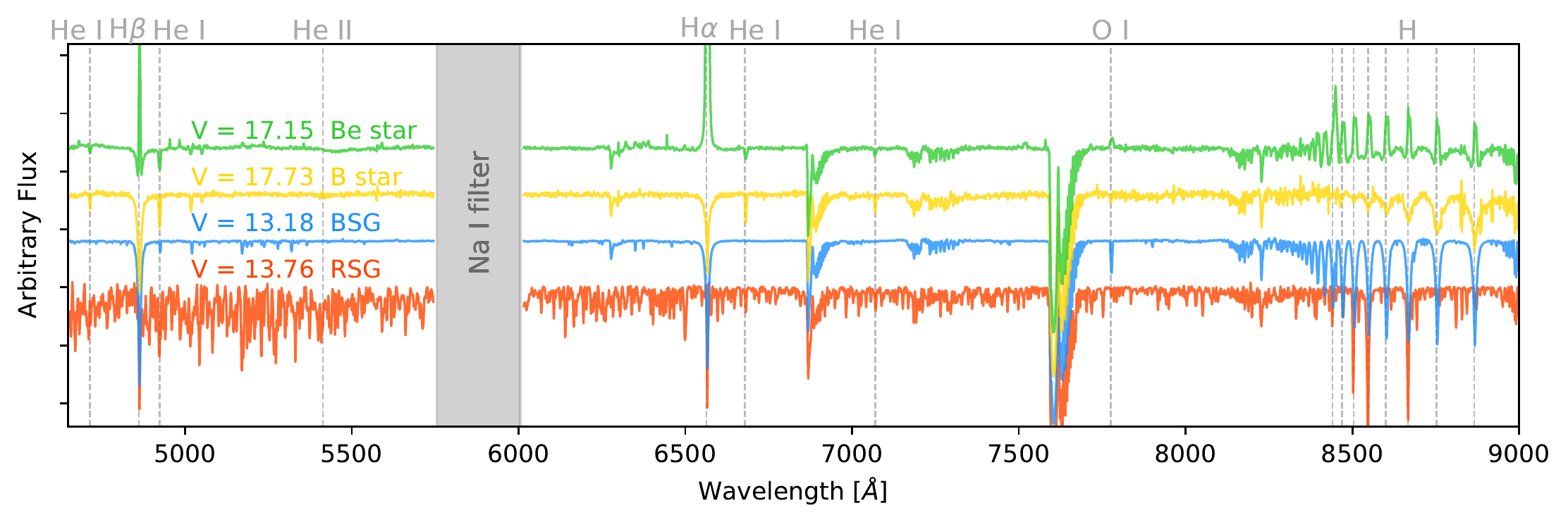}
  \caption{Example spectra for different stars. From top to bottom: the spectrum of a Be star (green), a B star (yellow), a BSG (blue) and a RSG (red). Our automated normalization mechanism was not applied to the RSG, it was normalized manually. The estimated $V$-band magnitude is given for each star. Important diagnostic lines are indicated. The gray shaded region indicates the wavelength region blocked by the sodium filter. The spectra are not corrected for tellurics.}
  \label{fig:example_specs}
\end{figure*}
After automatically extracting the spectra of the brightest stars we normalize them in an automatic manner. The normalization is performed for all spectra save for those of RSGs which require special treatment in account of the large number of spectral lines.\newline
To normalize the spectra, the continuum has to be determined and fitted with a polynomial. For this, spectral lines have to be identified and excluded from the fit. This continuum search is based on a two-step process adapted from \citet{Sana2013}. First, a minimum-maximum filter and a median filter are compared in a sliding window of a given size. If the value in the min-max filter exceeds the value of the median filter by a threshold value that depends on the S/N of the spectrum, a spectral line or cosmic hit is detected and the point is excluded from the fit. A polynomial is fit to this preliminary continuum. Points are excluded from the continuum points if they lie further away than a second threshold value from the polynomial fit, and a new polynomial is fit through the remaining continuum points. This procedure is repeated in an iterative manner.\newline
We consider the blue and the red part of the spectrum (i.e. wavelengths below and above the blocked laser-contaminated region between 5780 and 5990 $\AA$) independently. Furthermore, in the wavelength region dominated by the Paschen lines, continuum flux is determined in fixed wavelength windows.\newline
Normalized example spectra for a Be star, a B star, a RSG and a BSG are shown in Fig. \ref{fig:example_specs}. The estimated $V$-band magnitude is indicated in the plot. Using this method we achieve a normalization accuracy of around $2.5 \%$ for all the spectra. The average S/N is greater than 400 for the BSGs and varies between 300 for the brightest and 80 for the faintest B stars in our sample.

\section{Automated spectral classification}\label{Sec:specclass}
\subsection{Main sequence stars}\label{Sec:specclass_MS}
\begin{figure} \centering
\includegraphics[width=0.99\hsize]{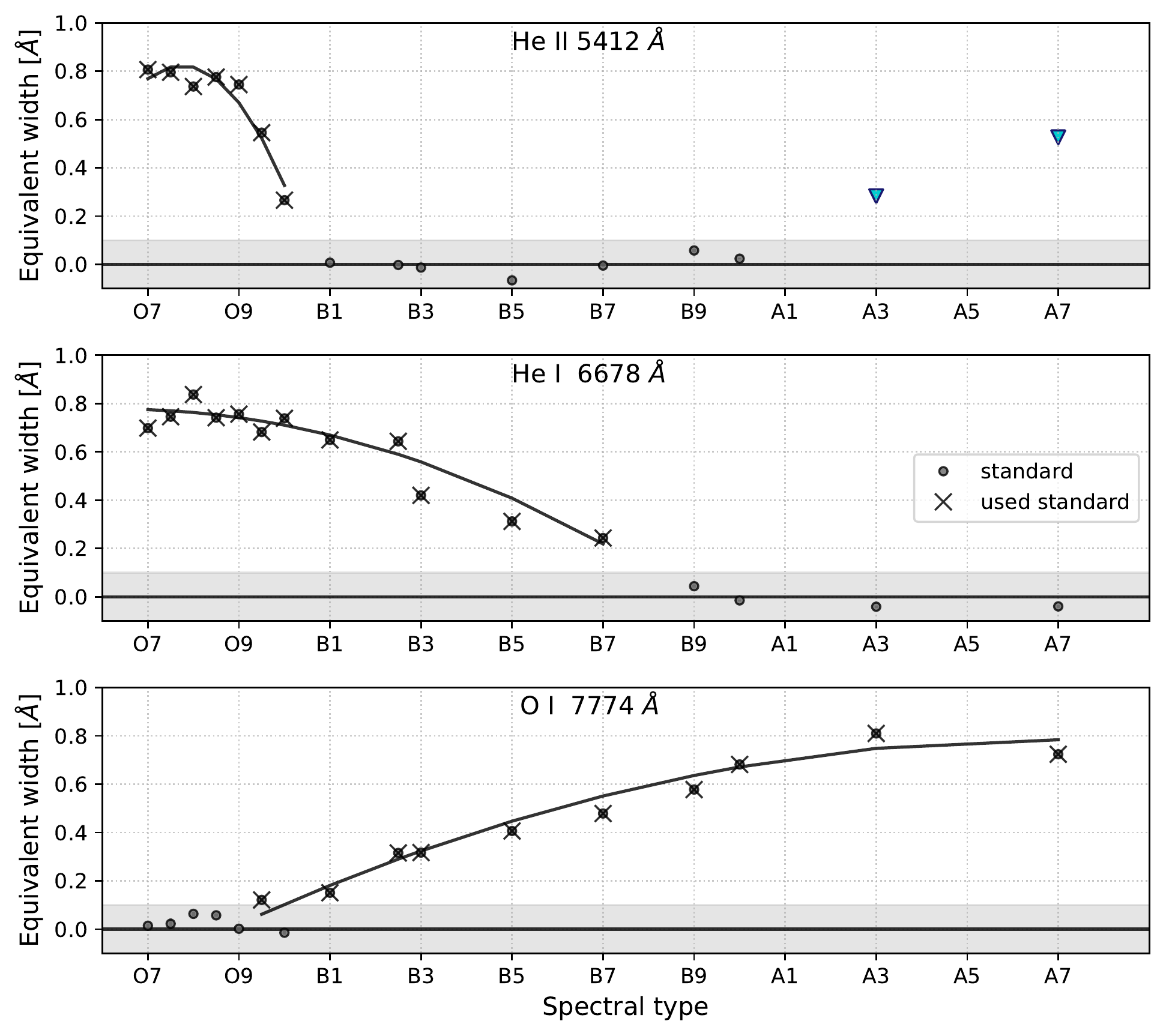}
\caption{From top to bottom: Correlation between equivalent width measured in the standard star spectra and their spectral type for He II $\lambda \,5412$, He I $\lambda \, 6678$ and the O I triplet at $\lambda \, 7774$. Black dots show the measured values while black crosses indicate the data points used to fit the polynomial. EW measurements below $0.1 \AA$ (shaded region around the black line at $0\,\AA$) do not allow a significant detection of spectral lines from noise in the continuum and are thus excluded from the fit. In the special case of the He II line the EWs at late spectral types (A5 and later), which are contaminated by metallic lines and thus not considered in the fit, are indicated by blue triangles.}
\label{fig:specew_empty}
\end{figure}

Commonly used spectral classification schemes are based on the blue part of the spectrum, i.e. on wavelengths below $4600\,\AA$ \citep{Conti1971, Walborn1990}. Unfortunately, this region is not covered by the MUSE extended setup (see Sec.\,\ref{Sec:MUSE}) and thus standard schemes cannot be applied to our case. Given the large amount of stellar spectra in our dataset we aim for an automated procedure that is reproducible and independent of human interaction. \newline
We develop a new classification scheme for OBA main-sequence stars in the red part of the spectrum that is based on the relation between spectral type and the equivalent width (EW) of spectral lines. Similar approaches were proposed by e.g. \citet{Kerton1999} or \citet{Kobulnicky2012} and applied to MUSE data by \citet{Zeidler2018} and \citet{McLeod2019}. While \citet{Kerton1999} focus on earlier spectral types between O5 and B0, \citet{Kobulnicky2012} make use of line ratios including the He I $\lambda\,5876$ line that is covered by the Na I filter in our data. Their classification can thus not be applied to our case. \newline
\begin{figure} \centering
  \includegraphics[width=1.05\hsize]{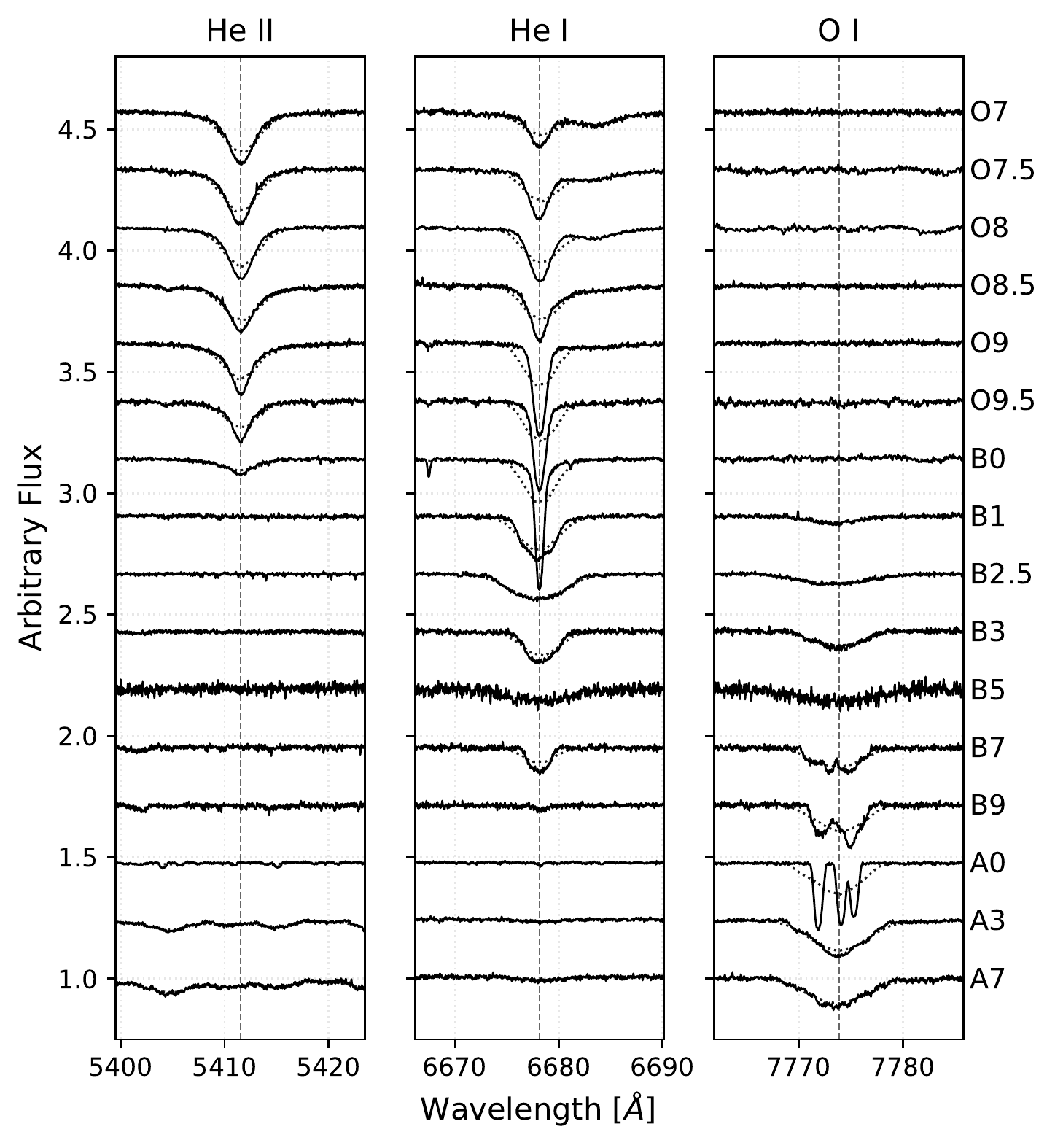}
  \caption{Atlas of spectral standards listed in Tab. \ref{tab:spec_calib} around the three lines of interest. The spectra are shifted horizontally according to the radial velocity and vertically for clarity. The black line shows the original spectra while the dotted line indicates the spectra degraded to MUSE resolution.}
  \label{fig:speccal}
\end{figure}
We select different diagnostic lines focusing in particular on He\,II $\lambda \, 5412$, He\,I $\lambda \, 6678$ and the O\,I triplet centered around $\lambda \, 7774$. EWs are not altered by stellar rotation or extinction and can be measured in a consistent manner at MUSE resolution.\newline
In order to calibrate the classification scheme we use the standard stars listed in \cite{Gray2009}. For all stars, observations obtained with the HERMES spectrograph mounted at the Mercator telescope in La Palma \citep{Raskin2011} are available. HERMES is a high-resolution spectrograph providing spectra with R = 85\,000. All classification spectra have a S/N\,$> 150$. Because of the lack of late-O and early-B spectral standard stars in \citet{Gray2009} we complement the list with the O star spectral atlas by Sana et al. (in prep.). The O star spectra used are observed with the Fibre-fed Extended Range Optical Spectrograph \citep[FEROS,][]{Kaufer1999} mounted at the 2.2m Max-Planck-ESO telescope in La Silla, Chile which has a resolving power of 48\,000. In a handful of cases both HERMES and FEROS observations are available for the same star. In these cases we use the higher resolution, higher S/N HERMES spectra. All spectra used for the calibration are shown in Fig. \ref{fig:speccal} as a sequence of spectral type.\newline
An overview of all calibration stars, their spectral types as listed in \cite{Gray2009} or Sana et al. (in prep.), and possible additional classifications are listed in Table \ref{tab:spec_calib}. Some stars show variable spectra. Several are single- or double-lined spectroscopic binaries or known pulsating stars. After downgrading the spectra to MUSE resolution we inspect the classification spectra to search for obvious contamination caused by the before mentioned variability. In almost all cases these are washed out due to the low resolution. Only $\nu$ Ori (B0V) shows a composite spectrum typical of a double-lined spectroscopic binary comprising an early type ($\mathrm{T_{eff}} \approx$ 20\,000 K) and a later type star ($\mathrm{T_{eff}} \leq$ 15\,000 K). It is thus not further considered in our classification. \newline
\begin{table*}
  \caption{OBA spectral standards from \citet{Gray2009} and Sana et al. (in prep.). The radial velocities (col. 3) and additional information (col. 4) are taken from references indicated in col. 5. Stars marked with a $\dagger$ are listed in Sana et al. (in prep.). The top part of the table gives main sequence standard stars (see Sec. \ref{Sec:specclass_MS}) while the bottom part gives giant and supergiant standards (see Sec. \ref{Sec:specclass_RSG}).}
  \begin{small}
  \begin{tabular}{llrlll} \hline \hline
    Classification & Identifier & v$_\mathrm{rad}$ & Additional information & References \\
     & & [km/s] & & \\ \hline
    O7V & 15 Mon & 22.0 & Oe star, SB2 (O7Ve + O9.5n) & \citet{Gies1993} \\
    O7.5Vz $\dagger$ & HD 152590 & 0.0 & eclipsing binary, SB2 & 
    \citet{Sota2014}, Sana et al. (in prep.) \\
    O8V $\dagger$ & HD 97848 & - 7.0 & - &  \citet{Walborn1982}, \citet{Sota2014}, Sana et al. (in prep.)\\
    O8.5V & HD 46149 & 45.9 & SB2 (O8.5V + B0-1V) & \citet{Mahy2009}, Sana et al. (in prep.) \\ 
    O9V & 10 Lac & -10.1 & - & \citet{Sota2011} \\ 
    O9.5V $\dagger$ & HD 46202 & 53.0 & & \citet{Sota2011}, Sana et al. (in prep.) \\
    B0.2V $\dagger$ & $\tau$ Sco & 2.0 & magnetic star & \citet{Donati2006} \\
    B1V $\dagger$ & $\omega$ Sco & - 4.4 & variable star of $\beta$ Cep type & \citet{Telting1998}, Sana et al. (in prep.) \\ 
    B2.5V(n) $\dagger$ & 22 Sco & 0.0 & - & \citet{Braganca2012}, Sana et al. (in prep.) \\ 
    B3V & $\eta$ Aur & 7.3 & - & \citet{Morgan1973} \\
    B5V & HD 36936 & 18.7 & - & \citet{Schild1971} \\
    B7V & HR 1029 & - 1.5 & pulsating variable star & \citet{Grenier1999, Szewczuk2015} \\
    B9V & HR 749A & 9.7 & - & \citet{Gray2009} \\
    A0Va & $\alpha$ Lyr & - 20.6 & variable star of $\delta$ Sct type & \citet{Gray2003, Bohm2012} \\
    A3V & $\beta$ Leo & - 0.2 & variable star of $\delta$ Sct type & \citet{VanBelle2009, Bartolini1981} \\
    A7V & HR 3974 & - 11.4 & variable star of $\delta$ Sct type & \citet{Gray2003, Rodriguez1991} \\\hline
    A0 Ia & HR 1040 & - 6.2 & evolved supergiant star & \citet{Bouw1981, Gray1987} \\
    A0 Ib & $\eta$ Leo & 1.4 & - & \citet{Cowley1969, Lobel1992} \\
    A0 III & $\alpha$ Dra & - 13.0 & SB1 & \citet{Cowley1969, Levato1972, Bischoff2017} \\ \hline
  \end{tabular}
  \end{small}
  \label{tab:spec_calib}
\end{table*}
For all the standard stars we measure the EW and the corresponding error $\sigma_\mathrm{EW}$ in a 16$\AA$-wide window $w$ around the center of the three lines mentioned above. The errors are estimated following \citet{Chalabaev1983}:
\begin{equation}
\sigma_\mathrm{EW} = \frac{\sqrt{2 \cdot w \cdot \Delta\lambda}}{\mathrm{S/N}}
\end{equation}
where the S/N is the signal-to-noise ratio of the continuum close to the considered line and $\Delta\lambda$ is the wavelength step (i.e. $1.25\,\AA$ for MUSE). We use these measured EWs to fit a second degree polynomial giving a relation between EW and spectral type.\newline
We repeat the measurements with degraded standard star spectra in which we adopt the resolution of MUSE and apply a typical S/N to the standard star data. While the relations themselves do not change, we use these measurements to investigate the limit for which genuine spectral lines can be distinguished from noise in the continuum. Given our data quality, this limit is $0.1\,\AA$. EW values are only considered for the relation fit if they are larger than this limit. Figure \ref{fig:specew_empty} shows the relations between EWs and spectral type for all three considered lines.\newline
A special case is the He II line at $5412\,\AA$. In late-type stars (i.e. A3 and later) metal lines start to dominate in the considered wavelength regime and especially the Fe I ($\lambda\,5410.9\,\AA$) and the Ne I line ($\lambda\,5412.6\AA$) contribute to EW measurements (marked with blue triangles Fig.\, \ref{fig:specew_empty}). We thus exclude these points from the relation fit at late spectral type. \newline
The derived relations between a fractional spectral type ($x_\mathrm{Spt}$) and the EW in $\AA$ for the considered lines are:
\begin{flalign}
x_\mathrm{Spt} = - \frac{0.4402 - \sqrt{0.3215 - 0.3905 \cdot \mathrm{EW_{5412}}} }{0.1952}; x_\mathrm{Spt} \in [-3, 0]\\
x_\mathrm{Spt} = - \frac{0.0360 - \sqrt{0.0152 - 0.0195 \cdot \mathrm{EW_{6678}}} }{0.0098}; x_\mathrm{Spt} \in [-3, 9]\\
x_\mathrm{Spt} = + \frac{0.0809 - \sqrt{0.0075 - 0.0096 \cdot \mathrm{EW_{7774}}} }{0.0048}; x_\mathrm{Spt} \in [0, 17]
\end{flalign}
in which we consider a linear translation of spectral type to $x_\mathrm{Spt}$ where B0 corresponds to $x_\mathrm{Spt} = 0$ and B9 to $x_\mathrm{Spt} = 9$.\newline
Figure \ref{fig:specew_empty} shows these results graphically: the differentiation between early B- and late O-type stars is per definition the presence of significant He\,II lines. The EW of the He\,II line at $5412\,\AA$ can thus be used as classifier for the presence of O stars in our sample. As He\,I lines are present in late O- to late B-type stars, our second relation can be extrapolated to spectral type B9 and thus be used for stars O7-B9. Finally, the O\, triplet at $7774\,\AA$ is present in spectral types later than B3 (up to A7). \newline
The derived relations are based on the standard star spectra of galactic stars and are not corrected for the effect of metallicity. As NGC~330 is in the SMC, i.e. at significantly lower metallicity, the O\,I line is weaker in our sample than predicted by the galactic standard stars. As the effect of metallicity is negligible for the He\,I and He\,II lines, we focus on those lines to determine spectral types in the MUSE data.\newline
\begin{figure} \centering
  \includegraphics[width=0.99\hsize]{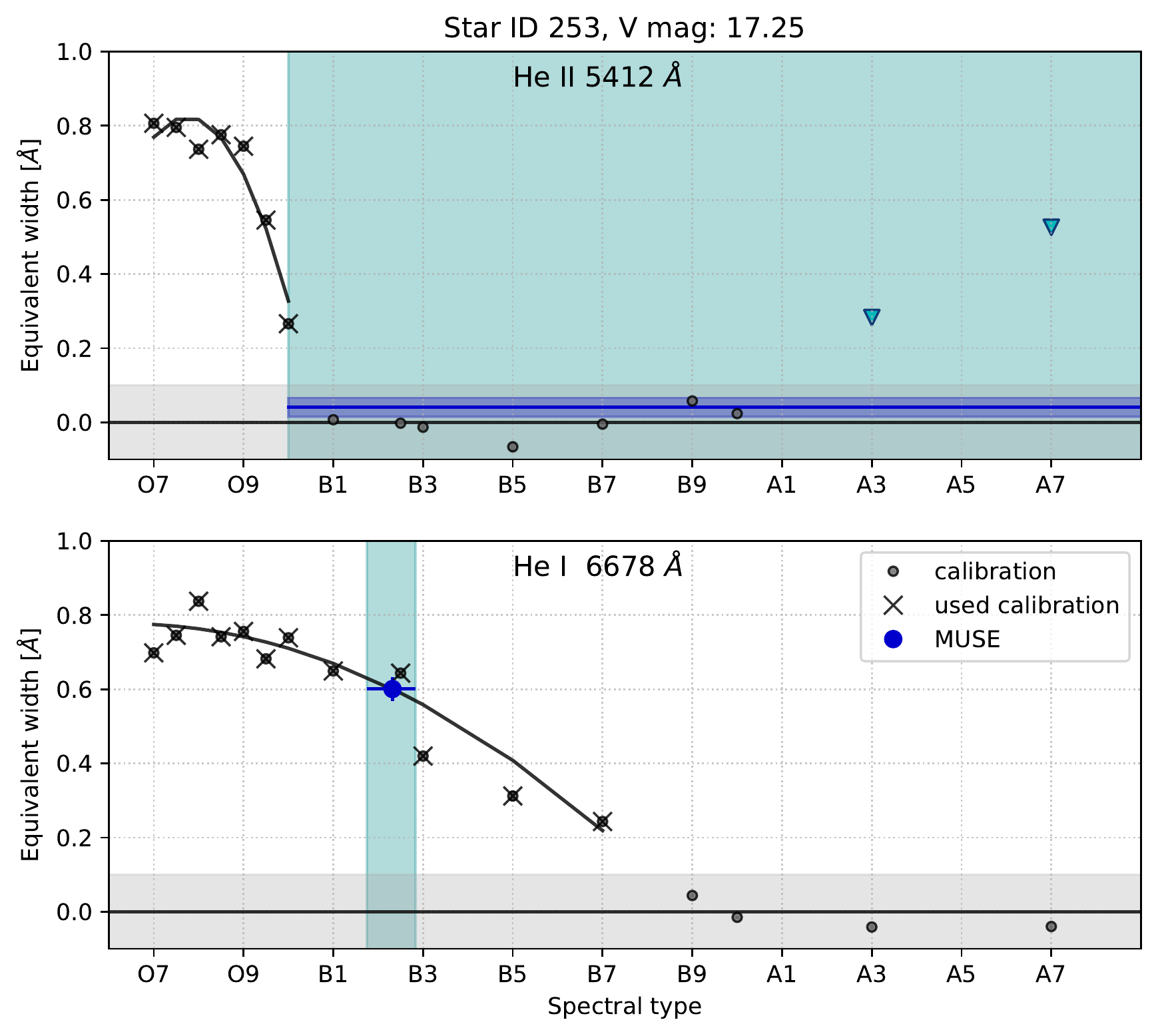}
  \caption{EW measurements for star \# 253 overplotted over the fit relations described in Equations (4) and (5). As mentioned above, we focus on the He lines to determine the spectral type. The blue shaded region indicates the allowed range of spectral types given the measured EWs with corresponding errors. Using this automated method, we determine a spectral type B2 with an accuracy of about one subtype. See also Fig.\,\ref{fig:ID253_eye}.}
  \label{fig:ID253_ews}
\end{figure}
\begin{figure} \centering
  \includegraphics[width=\hsize]{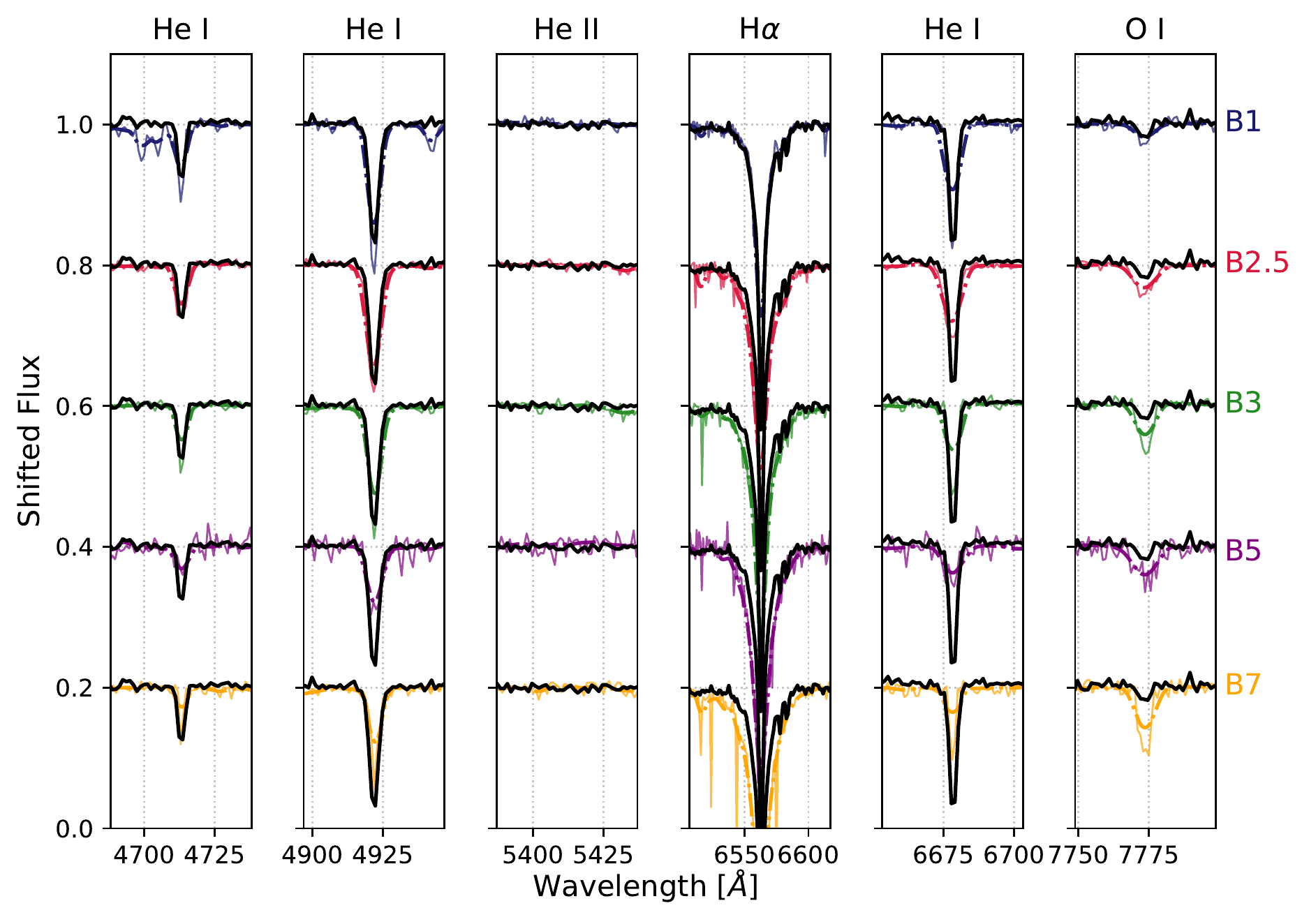}
  \caption{By-eye classification of star \#253. The stellar spectrum (in black) is compared to five standard stars, downgraded to MUSE resolution. The continuous lines show the original spectra while the dashed dotted lines show the spectra rotationally broadened by 200 km/s. By eye, the star would be classified as B1-B2 star, which is in agreement with the automatically derived spectral type of B2 $\pm$ one subtype. See also Fig. \ref{fig:ID253_ews}.}
  \label{fig:ID253_eye}
\end{figure}
The strength of the O\,I triplet at $\lambda \, 7774$ is sensitive to temperatures covered by later spectral types, i.e. especially to A-type stars where the He\,I  $\lambda \, 6678$ line is not present anymore. In order to be able to detect possible A-type stars we first classify all stars including the O\,I line. We do not detect any spectrum only showing O\,I lines and no He\,I, i.e. there are no A-type main-sequence stars in the sample. The only cases where O\,I is present in the spectra also show strong He\,I lines, a signature indicative of the composite spectra of spectroscopic binaries (see Sec. \ref{Sec:SBs}).\newline
The classification using the He\,II and He\,I lines is applied to all B-type main-sequence stars and yields fractional spectral types interpolated from the beforementioned relations. Taking into account the errors we achieve an accuracy of around one spectral subtype. An overview over the derived values is given in Appendix \ref{app:table_spectype}. An example diagnostic plot showing the measured EW for one star (\#253) and the derived spectral type is given in Fig. \ref{fig:ID253_ews}. For comparison reasons, Fig. \ref{fig:ID253_eye} shows the classification of the same star by eye. Both methods agree within their errors.\newline
Due to the difficulty to distinguish dwarf and giant luminosity classes given the low resolution of the MUSE data we derive the automated classification scheme for main-sequence stars only, implicitly assuming that all stars, but supergiants, are on the main sequence. We apply it to all sample stars except supergiants (that are several magnitudes brighter and clearly identifiable in the MUSE data, see Sect. \ref{Sec:specclass_RSG}). Besides RSGs and BSGs, the automatic approach to classification reveals its limitations for Be stars and spectroscopic binaries showing a composite spectrum. The identification and classification of these stars is described in the following sections.

\subsection{Red \& blue supergiants}\label{Sec:specclass_RSG}
\begin{figure} \centering
  \includegraphics[width=1.\hsize]{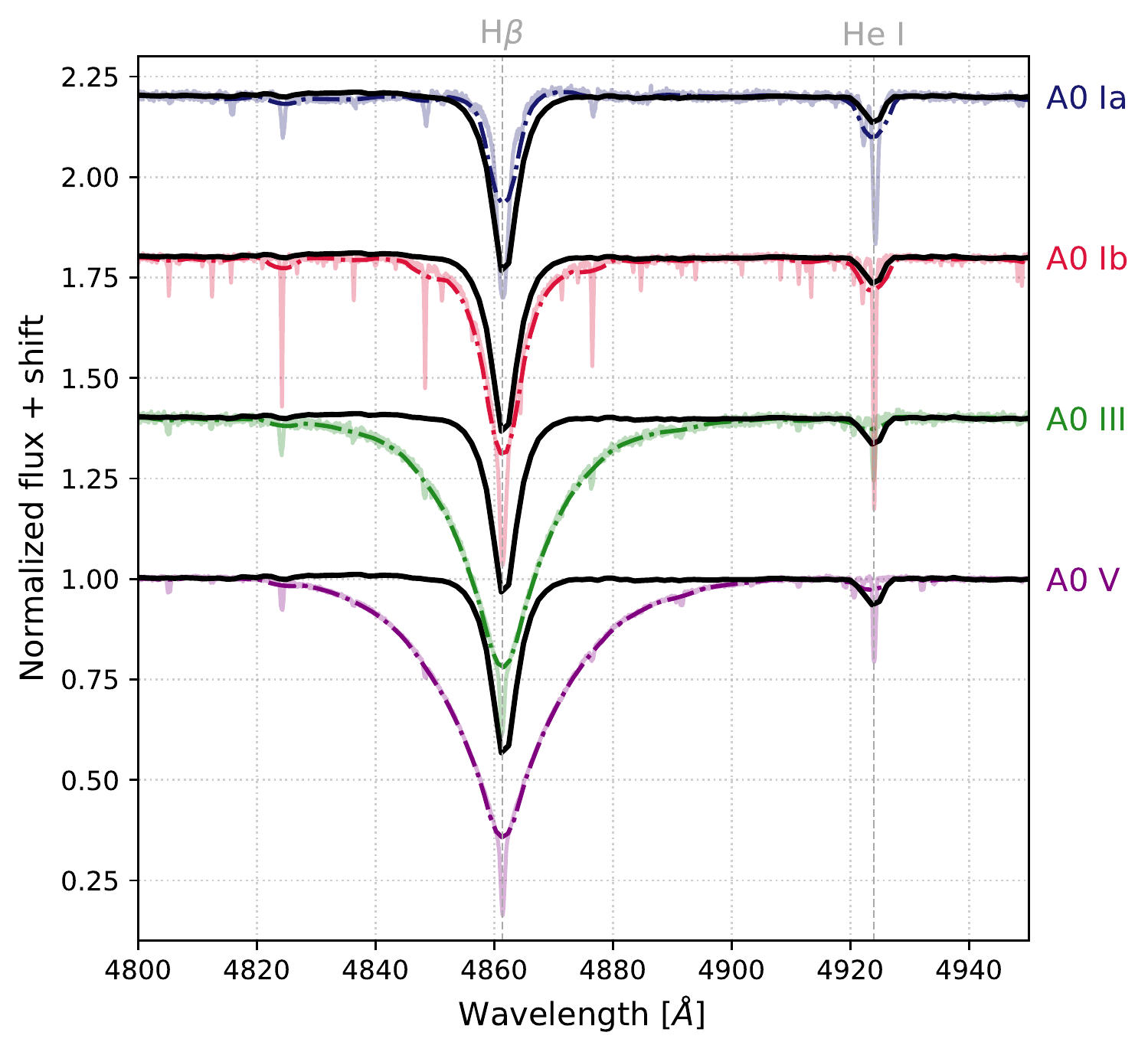}
  \caption{Spectral classification of an A-type supergiant (\#161). The MUSE spectrum (in black) is compared to four standard star spectra of spectral type A0\,Ia, A0\,Ib, A0\,III, and A0\,V. The continuous line shows the original spectra while the dashed dotted lines show the spectra rotationally broadened by 200 km/s. All standard star spectra are downgraded in resolution to match the MUSE data. We classify this star as early A-type star with luminosity class Ia-Ib.}
  \label{fig:supergiant}
\end{figure}

\begin{table*} \centering
  \caption{Overview of the parameters of the red and blue supergiants, both from this paper and from literature.}
  \begin{tabular}{llllllllll}\hline \hline
    \multicolumn{3}{c}{Identifier} & $\alpha$ & $\delta$ & V mag & \multicolumn{4}{c}{Spectral type} \\
    This work & A59 & R74 & J2000 & J2000 & R74 & this work & P2019 & FB80 & C85 \\ \hline

    416 & II-26 & A3 & 00:56:21.670 & -72:27:35.859 & 13.8 & - & G3 Ib & - & - \\
    437 & II-43 & A6 & 00:56:13.895 & -72:27:32.402 & 13.5 & - & K1 Iab & - & - \\
    356 & II-41 & A7 & 00:56:14.562 & -72:27:42.451 & 13.0 & - & G6 Ib & - & - \\
    297 & II-42 & A9 & 00:56:12.861 & -72:27:46.631 & 14.1 & - & K1 Ib & - & - \\
    042 & II-17 & A14 & 00:56:19.037 & -72:28:08.237 & 13.5 & - & K1 Ib & - & - \\
    420 & II-32 & A27 & 00:56:17.083 & -72:27:35.011 & 13.9 & - & K1 Ib & - & - \\
    285$^{\ast}$ & II-21 & A42 & 00:56:20.142 & -72:27:47.094 & 13.9 & - & K0 Ib-II & - & - \\
    195$^{\ast}$ & II-38 & A45 & 00:56:18.828 & -72:27:47.498 & 13.8 & - & G3.5 Ia-Iab & - & - \\
    237 & II-37 & A46 & 00:56:18.016 & -72:27:49.852 & 13.8 & - & K0 Ib & - & - \\
    279$^{\ast}$ & II-36 & A52 & 00:56:17.371 & -72:27:52.530 & 13.6 & - & K0 Ib & - & - \\
    183$^{\ast}$ & II-19 & A57 & 00:56:19.472 & -72:27:53.220 & 13.6 & - & G7 Ia-Iab & - & - \\ \hline
    210 & II-16 & A16 & 00:56:21.947 & -72:27:51.15 & 13.1 & early A Ib-III & - & B9 I & A2 I \\
    217 & II-20 & A41 & 00:56:20.186 & -72:27:50.99 & 13.2 & early A Ib & - & - & - \\ 
    288 & II-35 & A47 & 00:56:17.002 & -72:27:46.49 & 12.6 & early A Ia-Ib & - & B9 I & - \\ 
    296 & II-40 & A29 & 00:56:15.811 & -72:27:46.17 & 13.1 & early A Ib-III & - & - & - \\ 
    334 & II-23 & A19 & 00:56:21.022 & -72:27:43.92 & 13.1 & late A Ib & - & - & - \\ 
    397 & II-33 & A25 & 00:56:17.340 & -72:27:38.13 & 13.1 & early A Ib-III & - & A1 I & A0 I \\ \hline
  \end{tabular}
  \tablefoot{Literature identifications are from \citet[][A59]{Arp1959} and \citet[][R74]{Robertson1974}. Literature spectral types are from Patrick et al. (subm., P2019), \citet[][FB80]{Feast1980}, and \citet[][C85]{Carney1985}. $^{\ast}$ identifies potential Galactic contaminants based on Gaia data.}
  \label{tab:RSGBSG}
\end{table*}
Red and blue supergiants are identified based on their brightness, they are about several magnitudes brighter in the $V$ band than the main sequence stars. We find eleven bright red and six bright blue stars in the cluster core. Focusing on the bright red stars, Lee et al. (in prep.) check cluster membership based on Gaia DR2 proper motion and parallax measurements \citep{Collaboration2018}. They find that four stars have less reliable Gaia data and might be potential Galactic contaminants. Their radial velocities are, however, in good accordance with the SMC radial velocity. Given the large uncertainties in Gaia DR2 parallaxes for the bright blue stars we assess SMC membership by measuring their radial velocities, which are all in good agreement with the radial velocity of the SMC. In total, there are eleven RSGs and six BSGs in the cluster core. \newline
In order to verify that the stars are supergiants rather than giants we calculate their absolute $V$ magnitude using the distance modulus $m - M = 5\log d - 5 + A_V$, assuming a distance of $60$\,kpc and an extinction of $E(B-V)=0.08$ \citep[][D. Lennon, priv. comm.]{Keller1999}. As shown in Tab. \ref{tab:RSGBSG}, the $V$-band magnitudes of the RSGs and BSGs are between 12.6 and 14.1. This gives absolute magnitudes between -6.5 and -5.0 which is in the range of expected absolute magnitudes for bright A-type giants and A-type supergiants.\newline
The eleven RSGs in NGC 330 are studied in greater detail in Patrick et al. (subm.) who derive spectral types following the classification criteria and methodology detailed in \citet{Dorda2018}. We adopt their spectral type classification and include them in Tab.\,\ref{tab:RSGBSG}.\newline
For the classification of the six blue supergiants we use a second set of standard star spectra of different luminosity class from \citet{Gray2003} observed with the HERMES spectrograph. Standard star spectra are available for spectral types B5, A0, and A7 and include luminosity classes Ia, Ib, and III. \newline
When classifying the blue supergiants from the MUSE data, we first determine the spectral type based on He I lines and then estimate the luminosity class based on the width of the Balmer lines. As explained above, the strength of metallic lines can not be used for classification due to the lower metallicity of the SMC. Given the availability of the classification spectra we give a wide spectral type estimate of late-B, early-A and late-A.\newline
Five of the six blue supergiants are early-A stars while one is a late A-type star. While one star has luminosity class Ia-Ib, all the other supergiants are of luminosity class Ib-III. One example for the spectral classification of an A-type supergiant is shown in Fig. \ref{fig:supergiant}. \newline
Three of the A-type supergiants have already been studied in previous works by \citet{Feast1980} and \citet{Carney1985}. Their spectral types agree with the ones derived in this study within the errors. Two blue supergiants have not been mentioned in literature before. Table \ref{tab:RSGBSG} gives a summary of all the parameters of the RSGs and BSGs derived here compared to literature values.

\subsection{Be stars}\label{Sec:specclass_Be}
\begin{figure} \centering
  \includegraphics[width=0.99\hsize]{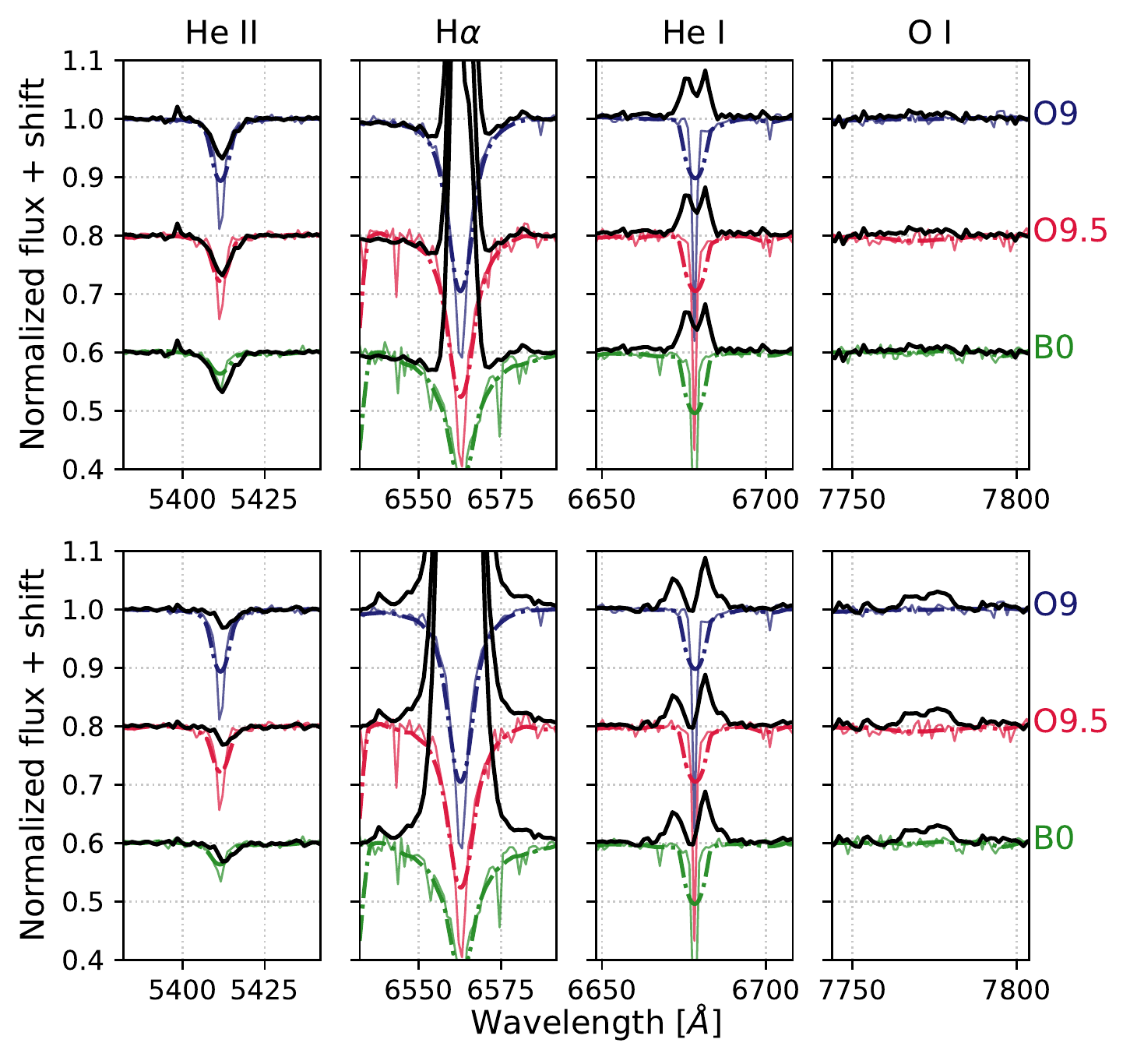}
  \caption{Spectral classification of the O9.5e star (\#147, upper panel) and the O9.5/B0e star (\#157, lower panel). The two stellar spectra (in black) are compared to three standard star spectra of spectral type O9, O9.5, and B0 in blue, red, and green (described in Sect. \ref{Sec:specclass_MS}), downgraded to MUSE resolution. The continuous line shows the original spectra while the dashed dotted line shows the spectra after applying rotational broadening by 200 km/s.}
  \label{fig:Oe}
\end{figure}

Classical Be stars are rapidly rotating B-type stars with Balmer lines as well as other spectral lines from e.g., O\,I and He\,I in emission, arising from a circumstellar decretion disk \citep{Rivinius2013}. Different possible origins for the rapid rotation of Be stars have been proposed. Be stars could be born as rapid rotators \citep{Bodenheimer1995, Martayan2007a}. They could also become rapid rotators during their MS evolution, either as single stars \citep{Ekstrom2008} or due to transfer of mass and angular momentum in a binary system \citep{Packet1981, Pols1991, Vanbeveren2017}. The strength of the emission lines is variable on timescales of months to years and strongly depends on the disk density and inclination angle \citep[see e.g.][]{Hanuschik1988}. Next to H$\alpha$, the O I $\lambda \, 7774$ line is most strongly affected by the infilling due to disk emission and often appears completely in emission. \newline
For this reason, besides automatically deriving a spectral type from the EW of the He I line, we compare the spectra of all Be stars to standard star spectra of B-type main-sequence stars by eye where a larger focus is put on all available He I lines. We achieve an accuracy with the visual inspection method of up to about 3 spectral subtypes. One example of this method of classification is shown in Appendix \ref{Sec:Bestar_specclass}. Both methods agree within the errors for a majority of the stars in the sample. In cases where the methods disagree we keep the spectral type that was derived through visual inspection as it allows us to takes into account more spectral lines.\newline
Both methods, however, are based on the strength of the He I lines which grow in strength with earlier spectral types. As they may suffer from infilling and thus appear less deep than they actually are, a systematic shift in classification towards later spectral types is possible. The estimated spectral types are thus a lower limit in terms of spectral type. \newline
In total we find Balmer line emission in 82 of the 251 MS stars in the sample. This gives a lower limit on the observed Be star fraction in our sample of $f_\mathrm{Be} = 32 \pm 3 \%$. While most of the emission line stars are of  mid- to late B-type , we find one O9.5eV and one O9.5/B0eV star in the sample. Both stars show clear absorption in the He II line at $\lambda \,5412$. Their spectra are shown in Fig. \ref{fig:Oe}.

\subsection{Spectroscopic binaries}\label{Sec:SBs}
\begin{figure} \centering
\includegraphics[width=0.99\hsize]{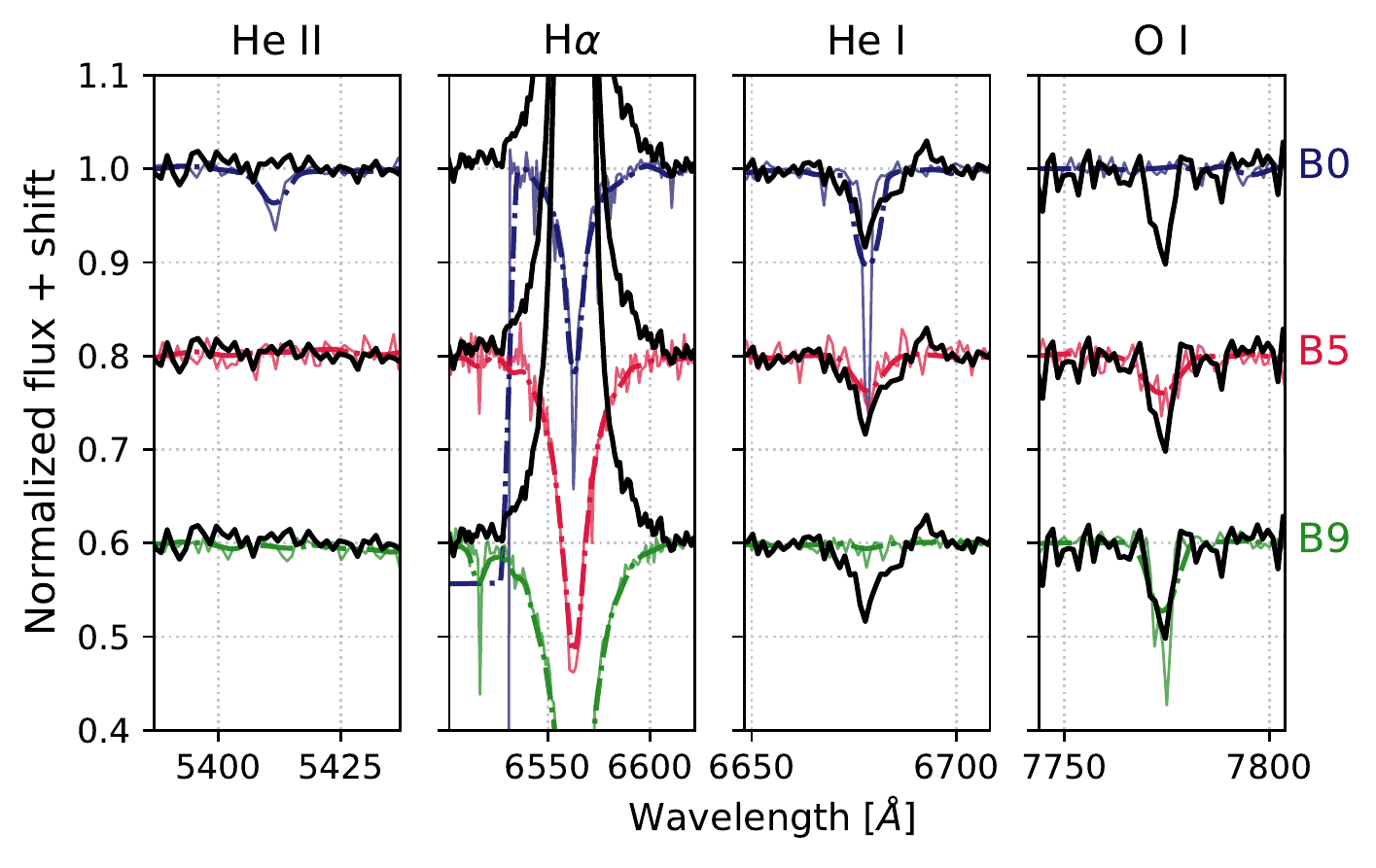}
\caption{Composite spectrum of the spectroscopic binary \#8. The spectrum of the star (in black) is compared to three standard star spectra of spectral type B0, B5, and B9 in blue, red, and green, respectively (described in Sect. \ref{Sec:specclass_MS}), downgraded in resolution to match the MUSE data. The continuous colored line shows the original spectra while the dashed dotted line shows the spectra after applying rotational broadening by 200 km/s.}
\label{fig:spec_binary}
\end{figure}

As we only consider one epoch in this study, we can not rely on radial velocity variations to detect spectroscopic binaries. Nevertheless we find indications for binarity in some of the spectra. For some stars we see strongly asymmetric lines made up of two line components of different radial velocity. Other spectra appear to be composite spectra showing both deep He I lines indicative of an early spectral type as well as deep O I lines characteristic for a later spectral type. \newline
Composite spectra could also occur when two sources are so close in the plane of the sky that their spectra are extracted together. The spectral extraction using PSF fitting based on the higher spatially resolved HST data as input catalog takes this into account. Other possible explanations of the observed line shapes could be, for example, different micro-/macroturbulence values that affect lines differentially. A higher oxygen abundance could lead to a sronger OI line. A more certain classification as spectroscopic binary will only be possible when periodic radial velocity shifts are detected.\newline
In total, we find 36 candidate spectroscopic binaries in our sample. If there is no false-positive due to the possible reasons mentioned above, this would imply a lower limit on the observed spectroscopic binary fraction of $f_\mathrm{binary} = 14 \pm 2 \%$. One example spectrum is shown in Fig. \ref{fig:spec_binary}. A more thorough investigation of the spectroscopic binaries will follow in paper II of this series where we will use all available 6 epochs to measure radial velocities for all stars.

\section{Stellar content of NGC 330}\label{Sec:results}
\subsection{Spectral types}
\begin{figure} \centering
\includegraphics[width=0.99\hsize]{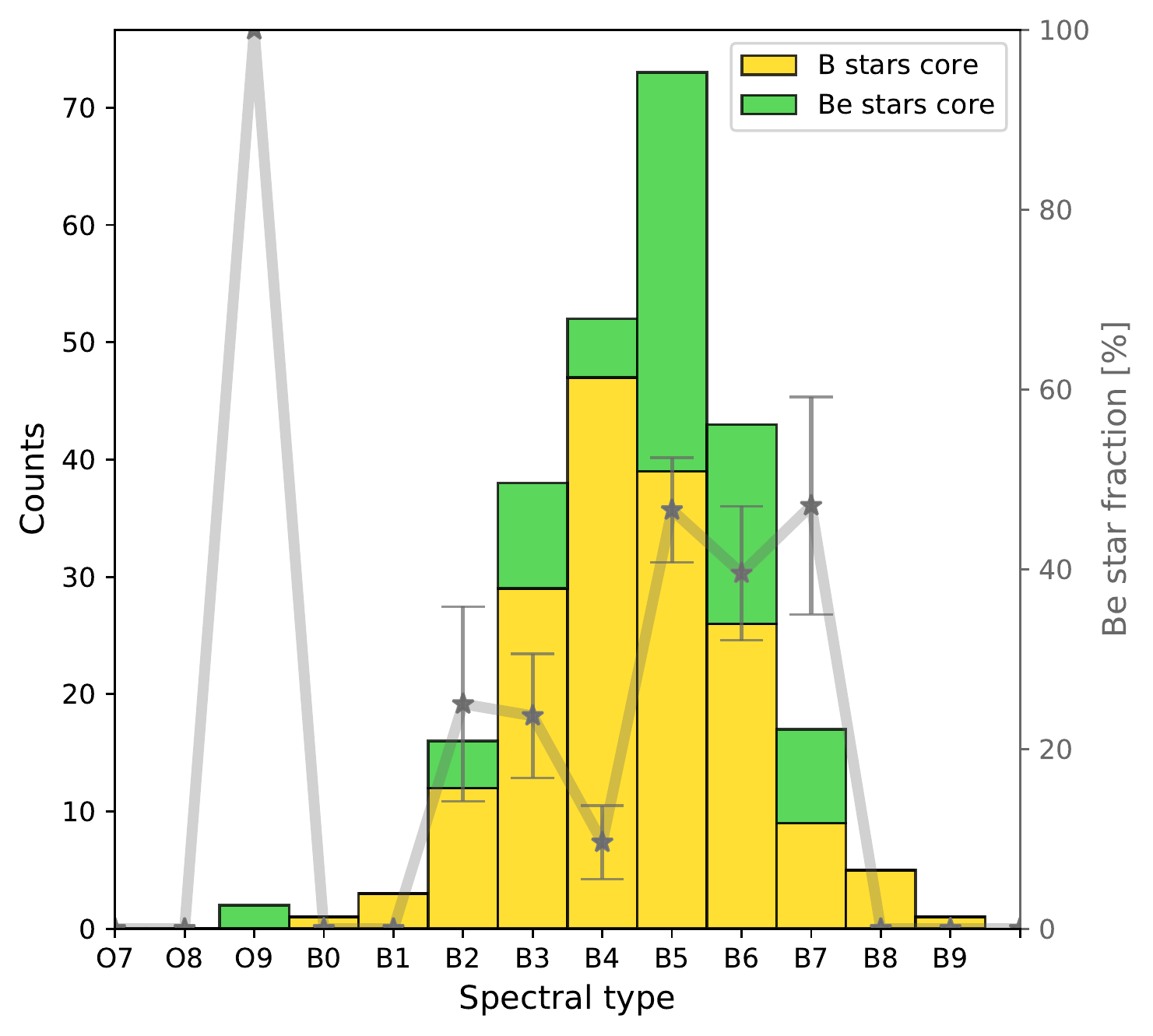}
\caption{Distribution of spectral types for main-sequence stars, where yellow indicates B stars and green indicates Be stars. The left axis shows the total number of stars while the right axis indicates the Be star fraction in each bin. The accuracy of spectral type determination is about one spectral subtype. Overplotted is the Be star fraction in each bin, for which we give counting errors except for cases where the Be star fraction is 0 or 100\%.}
\label{fig:histo_spectypes}
\end{figure}

\begin{figure} \centering
  \includegraphics[width=1.\hsize]{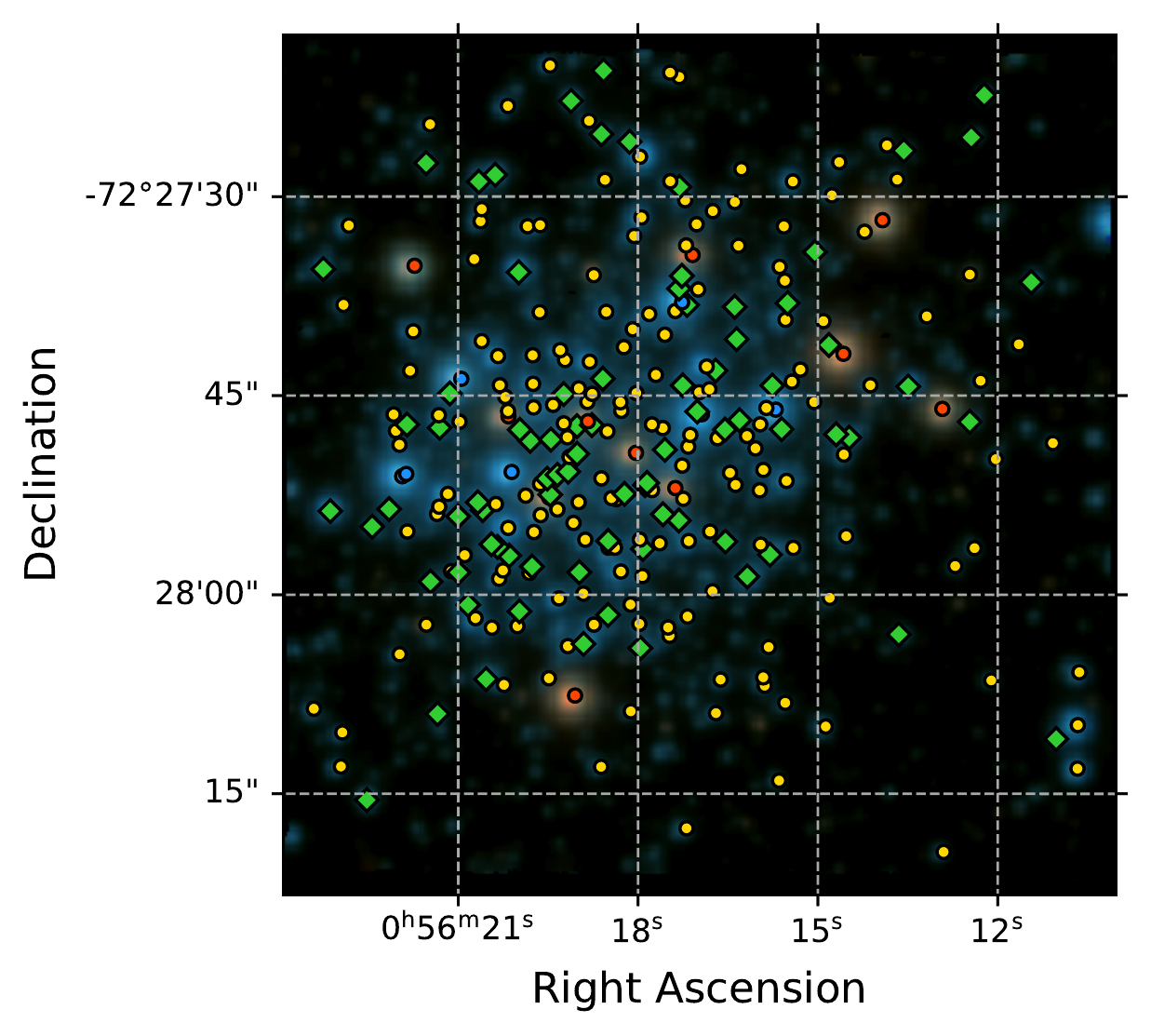}
  \caption{Stellar classification for all 275 stars with $V\,<\,18.5\,\mathrm{mag}$ overplotted over a MUSE RGB image created from the $B$, $V$, and $I$ filters. B stars are marked with yellow circles, Be stars with green diamonds, RSGs with red circles and BSGs with blue circles.}
  \label{fig:rgb_spectypes}
\end{figure}

Out of the 278 stars brighter than $V\,=\,18.5\,\mathrm{mag}$, there are 249 B type stars, two O type stars, eleven RSGs, and six A-type supergiants. For seven stars, the extraction of a sufficiently uncontaminated spectrum was not possible, and three stars appear to be foreground stars. 82 of the main sequence stars show Balmer line emission and are classified as Be stars. A distribution of spectral subtypes for B and Be stars is shown in Fig. \ref{fig:histo_spectypes}. Most of the stars are of spectral type B3 to B6. There are only a few stars with spectral type B2 and earlier and only one O-type and one O9/B0 star (both are in fact Oe/Be stars). The lack of later spectral types, i.e. of late-B and A type stars is probably due to the cut in brightness applied when extracting the spectra (i.e. only stars with $V\,<\,18.5$ are considered).\newline
The spatial distribution of all stars is shown in Fig. \ref{fig:rgb_spectypes}. While B-type stars seem to be evenly distributed, the Be stars seem to be clustered in the south-east part of the cluster core (i.e., the lower left corner of the image). Red and blue supergiants are mostly found in the inner region of the cluster.\newline
A handful of stars in our sample were studied by \citet{Grebel1996}, \citet{Evans2006}, and \citet{Martayan2007b}. The two stars studied by \citet{Grebel1996} have similar spectral types but different luminosity classes and were classified as Be stars by them while they show no signs of emission in our spectra (\#5 = Arp II-4 was as B2-3Ve star while we classify it as B3V star; \#461 = Arp II-31 was classified as B2-3 IIe star while we classify it as B5 V star). This could either be due to the variable nature of the Be star phenomenon, or it could be caused by H$\alpha$ contamination in the spectra used by \citet{Grebel1996}.\newline
Our derived spectral types for the two stars scrutinized by Evans et al. (\#011 = NGC 330-036 and \#253 = NGC 330-095), are within the error bars: we classify \#011 as B3 while it is listed as B2 II star in \citet{Evans2006}, and \#253 as B2, which by them is listed as B3 III star.\newline
Furthermore, there are six stars in common with the sample of \citet{Martayan2007b} which are classified as early Be stars (B1e-B3e). For those stars (\#53, \#88, \#140, \#166, \#456 and \#460) we find a systematically later spectral type between B3e and B6e. This systematic difference could be due to their different classification methods: while we determine spectral types based on the EW of diagnostic spectral lines and by visual comparison to standard star spectra, Martayan et al. estimate $T_\mathrm{eff}$ and $\log\,g$ from the spectra and transcribe these into spectral types and luminosity classes adopting the galactic calibrations by \citet{Gray1994} for B stars, and \citet{Zorec2005} for Be stars.

\subsection{Color Magnitude Diagram}\label{Sec:cmd}
\begin{figure} \centering
\includegraphics[width=0.99\hsize]{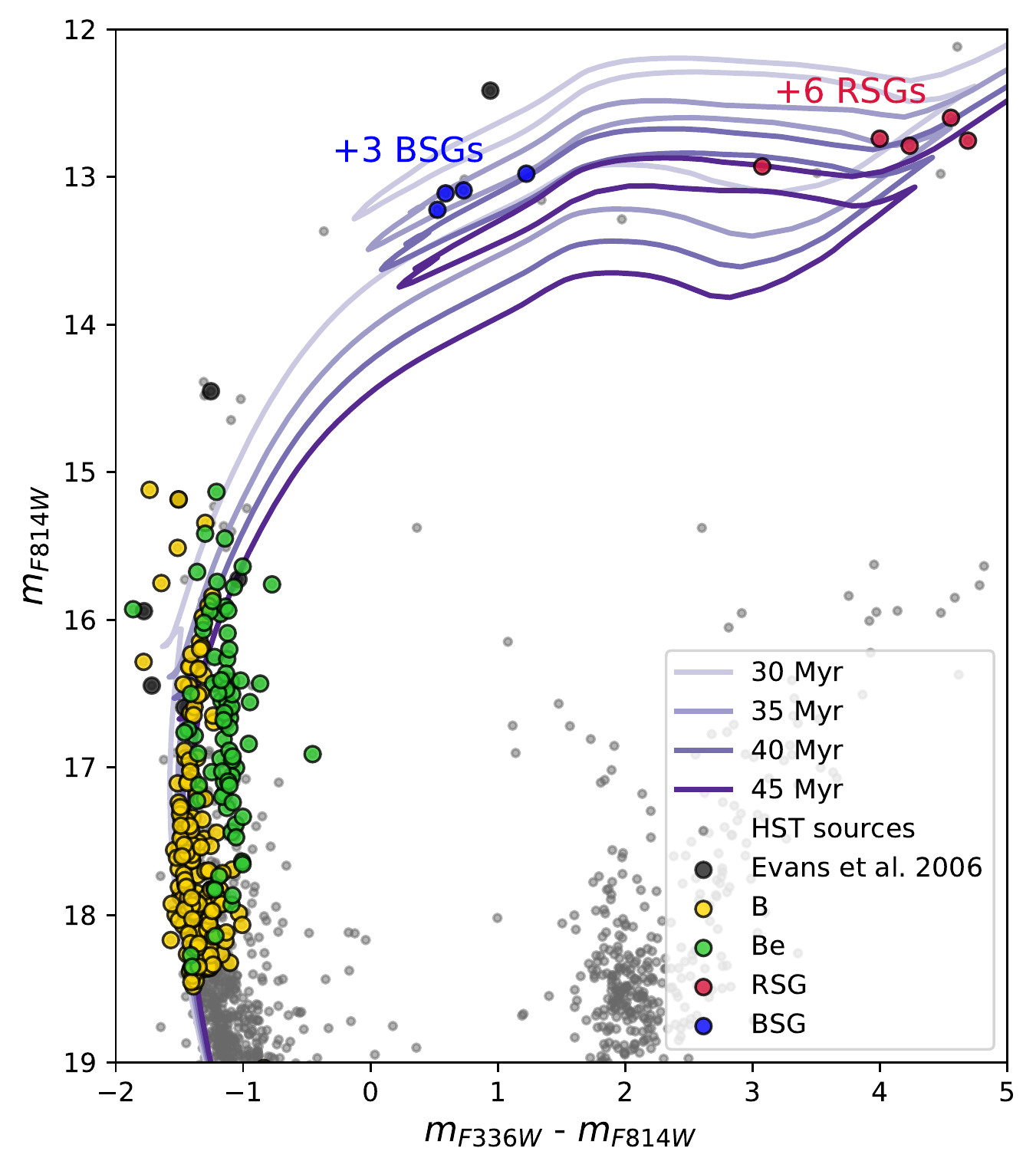}
\caption{Reconstructed CMD from \citet{Milone2018} where their complete input catalog containing field and cluster stars (see their Fig. 2) is shown in grey. The sources with MUSE spectra available are colored in terms of the spectral type we derived from the MUSE data, i.e. Be stars in green, B stars in yellow, RSGs in red and BSGs in blue. The eight sources studied by \citet{Evans2006} included in the HST data are indicated with black circles. Overplotted are single-star non-rotating Padova evolutionary tracks for ages between 30 and 45 Myr. Three A-type supergiants and six RSGs are missing in the HST catalog and thus can not be included in this diagram.}
\label{fig:CMD}
\end{figure}

We use the HST photometry published in \citet{Milone2018}\footnote{online available under http://progetti.dfa.unipd.it/GALFOR/} to reconstruct their color magnitude diagram (CMD) from the HST F336W and F814W bands (see Fig.\,\ref{fig:CMD}) and cross-correlate it with our MUSE data to include the spectral type information we obtain from the spectra. The HST observations contain around 20\,000 sources covering a significantly larger area than the MUSE FoV. Apart from NGC\,330 cluster members, the data probably contain several field stars as well as foreground stars (i.e. in lower right part of the CMD). Several of our brightest stars, i.e. two A-type supergiants and six RSGs are missing in the HST catalog. We can thus not include them in the CMD.\newline
We confirm the findings of \citet{Milone2018} who report a high fraction of stars with H$\alpha$ emission which they interpret as Be stars (taking into account narrow-band F656N observations). Furthermore they find a split main sequence near the cluster turnoff which they understand as signature of multiple stellar populations in the cluster. Using the MUSE spectra we confirm that the red MS consists of Be stars with H$\alpha$ emission.
This could either be due to the fact that the Be star phenonemenon preferentially occurs at or close to the terminal age main sequence \citep{Martayan2010}. It could also be a temperature effect due to the rapid rotation of these stars. According to \citet{Townsend2004} rapidly rotating stars may indeed appear cooler, translating into a classification of up to two spectral sub-types later. This effect can also be seen in Fig. \ref{fig:histo_spectypes}. \newline
Following \citet{Milone2018} we overplot single-star non-rotating Padova isochrones \citep{Bressan2012, Chen2014, Chen2015, Tang2014, Marigo2017, Pastorelli2019}\footnote{obtained from http://stev.oapd.inaf.it/cgi-bin/cmd}. We adopt the standard values for the SMC, i.e., the distance of $d=60\,\mathrm{kpc}$
\citep{Harries2003, Hilditch2005, Deb2010}, the metallicity $Z=0.002$ \citep{Brott2011}, the foreground extinction $E(B-V)=0.08$ \citep[][D. Lennon, priv. comm.]{Keller1999}, the SMC extinction law by \citet{Gordon2003}, and consider isochrones for ages between 30 and 45 Myr. While we find a good agreement of the isochrones of 35 and 40 Myr with the uppermost part of the main sequence, none of the isochrones reproduces the fainter end of the main sequence (stars fainter than $F814W = 20$,
not included in Fig.~\ref{fig:CMD} but see Appendix \ref{Sec:appendixCMD} for a zoomed-out version). They are systematically too faint by $\approx 0.3\,\mathrm{mag}$.\newline
This effect can also be seen in the CMDs showed in \citet{Milone2018} and \citet{Sirianni2002}. While \citet{Sirianni2002} use different HST data and different isochrones from \citet{Schaerer1993}, we use the data from \citet{Milone2018} and the same Padova isochrones with slightly different input values (i.e., they adopt $d = 57.5\,\mathrm{kpc}$, $E(B-V) = 0.11$, and a galactic extinction law). Trying to reproduce their findings using the same input values we find that the shape of the isochrones is slightly different. This might be due to updates in the codes used to derive the isochrones (several updates were released in 2019, including a new version of the code producing isochrones, \citet{Marigo2017}).\newline
When fitting the isochrone we thus focus on the upper part of the B-star MS which is best reproduced by the isochrones at 35 and 40 Myr. The isochrones at 30 and 45 Myr do not fit the top part of the stars (see also Fig.~\ref{fig:CMD_zoomout}). This leads to an age estimate of $35 - 40 $ Myr. We note that the isochrones reproduce the position of the RSGs and BSGs in the CMD very well. \newline
Patrick et al. (subm.) estimate the age of NGC~330 based on the population of RSGs \citep[see also][]{Britavskiy2019}. They find a cluster age of 45\,$\pm$5\,Myr, which is in good agreement with what is presented here, which is not surprising given the how well the isochrone fitting method reproduces the location of the RSGs in Fig. \ref{fig:CMD}.\newline
We find the cluster turn-off to be at $F814W \approx 16.5\,\mathrm{mag}$ which corresponds to $M\approx 7.5\,M_{\odot}$, which is just below the limiting mass for neutron star formation. This indicates that all currently produced binary Be stars may remain in the cluster, as their companions are likely white dwarfs.\newline
Several stars in the CMD are brighter and hotter than the cluster turn-off, i.e. they could be blue stragglers. Half of them are B stars while the other half are Be stars. These stars are expected to have gained mass in previous binary interactions. They are thus some of the best candidates for BiPs.

\subsection{Be stars}
\begin{figure} \centering
\includegraphics[width=0.99\hsize]{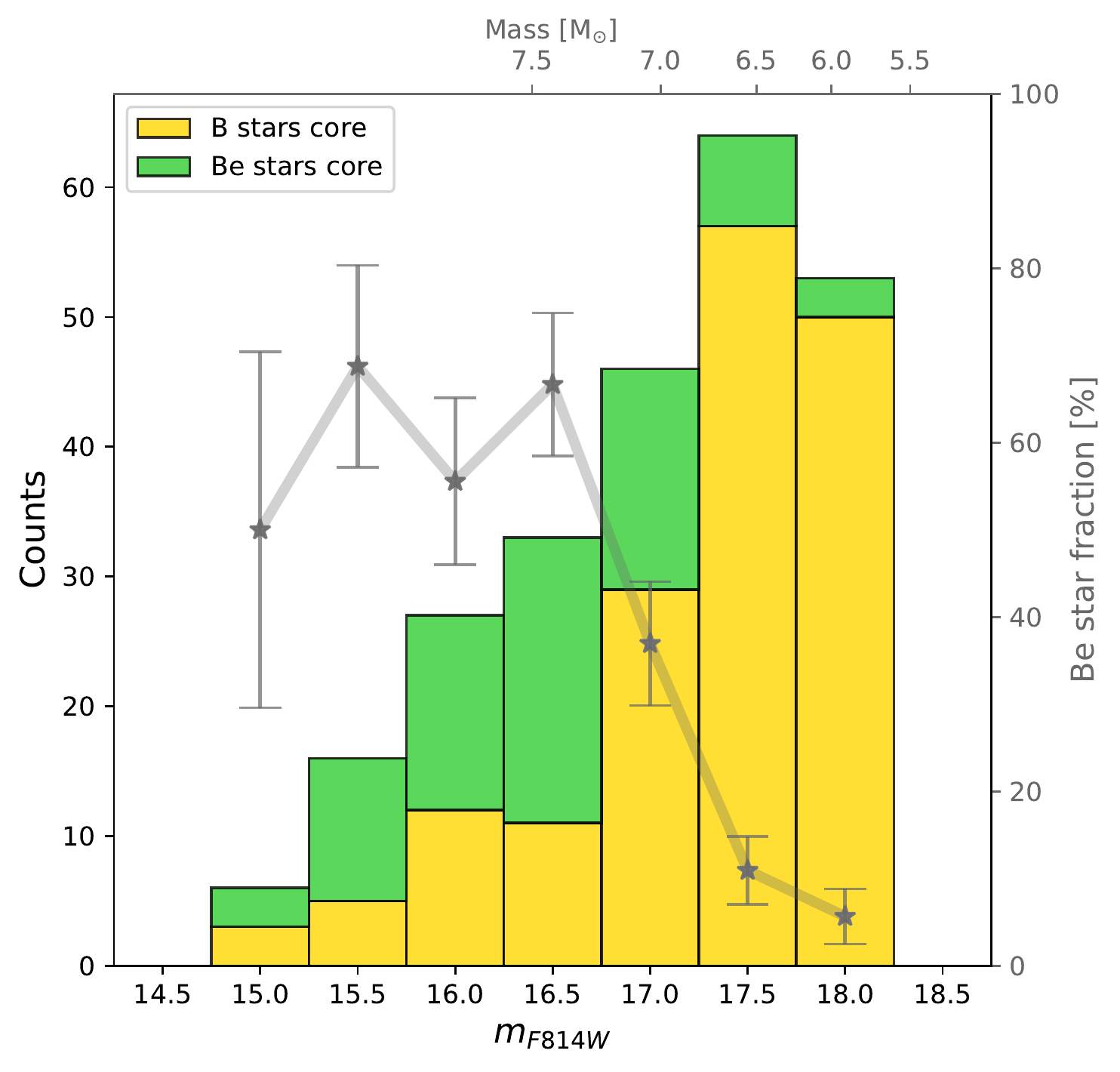}
\caption{Distribution of B and Be stars (in green and yellow, respectively) over F814 magnitude taken from \citet{Milone2018} where the left axis shows the total number of stars and the right axis indicates the Be star fraction in each bin. The top axis indicates the conversion from F814W magnitude to mass taken from the Padova isochrones. Stars brighter than $F814W=16.5$ are above the cluster turn-off where their mass remains mostly constant while they increase in brightness.}
\label{fig:Befrac_mag}
\end{figure}

Among the 251 B- and O-type stars there are 80 Be stars as well as two Oe stars, indicating a lower limit on the Be star fraction of $f_\mathrm{Be} = 32\pm$3\%. This fraction is high compared to typical Be star fractions of around 20\% \citep{Zorec1997} measured in field or cluster stars in the Milky Way. The latter fraction is determined using several years of observations of bright galactic stars. Due to the variability of Be stars and the disappearance of the emission lines on timescales of months or years \citep{Rivinius2013}, our fraction based on one observational epoch is a lower limit on stars that display the Be phenomenon.\newline
Several previous studies derive high Be star fractions (of the order of 50\%) in NGC 330 \citep{Feast1972, Mazzali1996, Keller1998}. They are, however, based on small sample sizes and focus on stars in the outskirts of the cluster. \citet{Evans2006} study a sample of 120 stars in the outskirts of NGC 330 (see Fig. \ref{fig:rgb}) and find a Be fraction of $f_\mathrm{Be} = 23 \pm 4$ \%. Our detected Be star fraction is significantly higher than the one measured by \cite{Evans2006}.\newline
Interestingly, the two earliest-type stars in the sample have H$\alpha$, He\,I, and O\,I emission while they also show He\,II absorption. They are classified as Oe/Be stars (see Sect. \ref{Sec:specclass_Be}). Using the CMD, we identify them as blue stragglers, i.e. BiP candidates. \newline
As shown by the Be star fraction in Fig. \ref{fig:histo_spectypes} the spectral type distribution of Be stars seems to be bimodal, with a shift to later spectral types. As explained in Sect. \ref{Sec:specclass_Be}, this could be due to the infilling of spectral lines by the emission component, which in particular leads to weaker He I lines and thus a classification into a later spectral type. Additionally, rapidly rotating stars are likely to appear cooler and would subsequently be classified up to two spectral types later than a non-rotating star \citep{Townsend2004}. \newline
In Fig. \ref{fig:Befrac_mag} we instead consider the fraction of Be stars as a function of HST magnitude (F814W) as it is less sensitive to these effects, assuming the population is coeval. It shows that the Be star fraction is between 50 and 75\% for stars brighter than 17.0\,mag in $F814W$ (corresponding to $7.0\,M_{\odot}$, according to Padova isochrones) and rapidly drops below 20\% for stars fainter than $17\,\mathrm{mag}$ in $F814W$.
This drop corresponds to the cluster turn-off at $\mathrm{F_{814W}} = 16.5\,\mathrm{mag}$ (i.e. $M\approx 7.5\,M_{\odot}$). We quantitatively estimate the effect of emission in the Paschen series on $F814W$ photometry from the MUSE spectra. We find that even the strongest Paschen emission series amounts to only $\approx 0.01\,\mathrm{mag}$ in the $F814W$ band, and hence is negligible. We furthermore see a similar trend when considering $F336W$ magnitudes instead of $F814W$ magnitudes in the histogram.\newline
The lack of fainter Be stars could indicate the cut-off luminosity (or mass) for which B-type stars do not appear as Be stars anymore. This could either be due to the fact that these stars do not develop a disk, or to fact that the ionizing flux emitted by these fainter stars is not high enough to ionize the Be star disk \citep{Bastian2017}.\newline
A high-mass X-ray binary ([SG2005] SMC38) is reported within the MUSE FoV \citep{Shtykovskiy2005, Sturm2013, Haberl2016}. Giving the crowding of the region and the uncertainty in the position of the source of 0.7" an unambiguous assignment to one of our sample stars is not possible. The two closest stars (i.e. within the errors of the XMM position) are \#68 and \#80, a B5Ve and a B4V star, respectively. In both cases, and given the position of the system in the cluster core, the massive but hidden companion may be a black hole. In the case of a neutron star, it is likely that the system would have been ejected in the supernova explosion. Given the distance of the SMC, even at low kick velocities of $\mathrm{v}_\mathrm{kick} \leq 2 \mathrm{km/s}$, the system would have moved 20\,pc within 10\,Myr, i.e. out of the MUSE FoV. A more in-depth analysis of this object will follow in a subsequent paper of this series.

\subsection{Estimate of the total mass of NGC 330}
In order to estimate the total mass of NGC 330 we make use of the HST data to determine a scaling factor for the initial mass function (IMF) to the local star density of NGC\,330. For this we need an estimate of the total stellar mass in a certain mass bin.\newline
Following \citet{Milone2018} we define the cluster core to have a radius of 24" and only consider stars within that distance from the center. As the HST data cover a wider region than the MUSE data we take a less crowded region towards the edge of the HST data in order to estimate the contamination by field stars. \newline
From the isochrone fit to the blue main sequence in the CMD (see Fig.\, \ref{fig:CMD}) we count the number of stars between 3 and 5 $M_{\odot}$ in the cluster as well as in the field region. We find that there are 412 between 3 and 5 $M_{\odot}$ , which corresponds to a total mass of 1650\,$M_{\odot}$ in this mass bin. In comparison, the mass of the field population in that same mass bin, which we subtract from the cluster mass, is only 94\,$M_{\odot}$. Comparing this to the integrated IMF in the same mass bin gives us a scaling factor that scales the IMF to the local star density of NGC\,330.\newline
Assuming a Kroupa IMF \citep{Kroupa2001} and the occurence of stars with masses between 0.01 and 100 $M_{\odot}$ we derive a total mass of the cluster of $M_\mathrm{tot,NGC330} = 88 \times 10^3\,M_{\odot}$.\newline
In order to estimate an error on the total cluster mass we inspect the sensitivity of the estimate to the assumptions we made. The total cluster mass increases by 5\% if we do not subtract the field population when estimating the mass between 3 and 5 $M_{\odot}$. While the mass estimate is not altered significantly by assuming a different high-mass end of the IMF, it depends strongly on the low-mass end: Taking a low-mass end of 0.08\,$M_{\odot}$ (instead of 0.01\,$M_{\odot}$) the total cluster mass is 4\% lower. The assumption that is affecting the estimate the most is, however, the assumed cluster radius. Increasing and decreasing the radius by 25\% results in a total mass of $M_\mathrm{tot,NGC330} = 105 \times 10^3 \,M_{\odot}$ and $M_\mathrm{tot,NGC330} = 70 \times 10^3 \,M_{\odot}$, respectively. We thus take this as the dominating uncertainty which results in a total cluster mass of $M_\mathrm{tot,NGC330} = 88^{+17}_{-18} \times 10^3 \,M_{\odot}$. \newline
\citet{Mackey2002} estimate a mass of $38 \times 10^3 \,M_{\odot}$ based on the surface brightness profile derived from archival HST data. \citet{McLaughlin2005} find a similar value of $36 \times 10^3 \,M_{\odot}$. As an order of magnitude estimate, we agree with their mass estimate within a factor of two. Patrick et al. (subm.) estimate the dynamical mass of NGC~330 based on the velocity dispersion of several RSGs and find a dynamical mass
of $M_\mathrm{dyn} = 158^{+76}_{-51} \times 10^3 \,M_{\odot}$, which is significantly larger than our estimate. The authors comment that this could be due to the effect of binarity on the determination of $M_\mathrm{dyn}$.

\subsection{Inner regions vs. outskirts}
\begin{figure} \centering
\includegraphics[width=0.99\hsize]{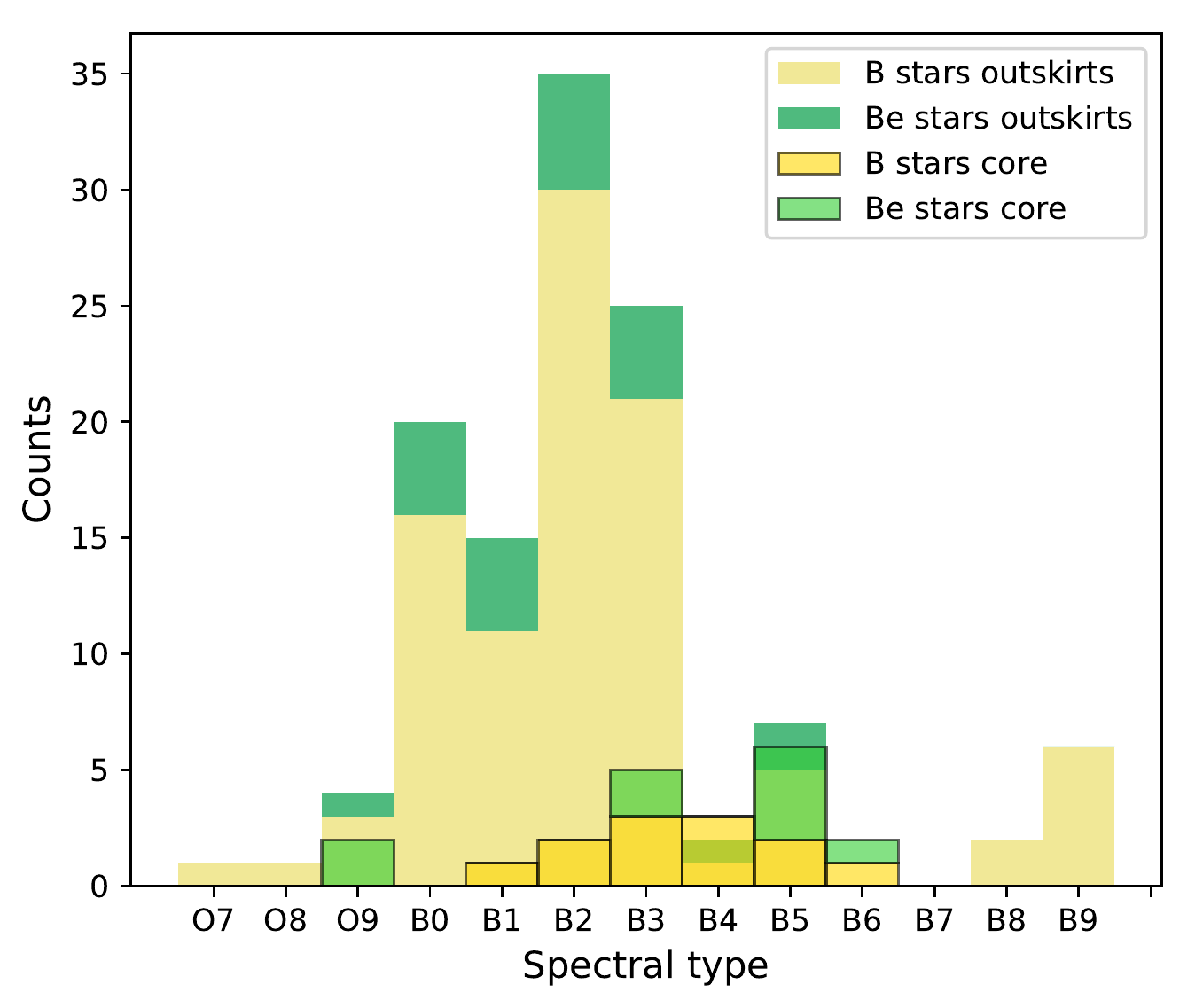}
\caption{Distribution of spectral types from \citet[][outskirts]{Evans2006} compared to the MUSE data (see Fig. \ref{fig:histo_spectypes}) of the core. As there are no stars fainter than $V\,=\,17$ in the sample of Evans et al. we cut our sample at this brightness to facilitate the comparison.}
\label{fig:histo_spectypes_evans}
\end{figure}

Figure \ref{fig:histo_spectypes_evans} compares the spectral type distribution of \citet{Evans2006} with a subsample of this work adopting the same magnitude cutoff, i.e. $V<17\,\mathrm{mag}$. This only leaves 22 main-sequence stars in our sample. It can be seen that there are significantly more early-type stars in the sample of \citet{Evans2006}. This trend persists when we only include stars from \citet{Evans2006} that are closer than a certain distance from the cluster core, reducing the probability of including field stars. This could be due to a systematic difference in spectral typing, although this is not supported by the two stars that are in common between the two samples (see Sec. \ref{Sec:results}). It could also indicate that the stars in the sample of Evans et al. are younger and more massive than the core population.\newline
Furthermore we find a significantly higher Be star fraction in the core compared to the outskirts. Restricting our sample to stars brighter than $V\,=\,17$, which is the brightness limit of Evans et al., we find a Be star fraction of $f_\mathrm{Be} = 46 \pm 10$ \%, while the fraction detected by Evans. et al. is $f_\mathrm{Be} = 23 \pm 4$ \%.\newline
Another significant difference between the core and the outskirts is the different binary fraction. Based on observations spread over only 10 days, \citet{Evans2006} report a particularly low observed binary fraction of only $(4\,\pm\,2)\%$. In this study, based on a single epoch, we find a lower limit on the observed binary fraction of $(14\,\pm\,2)\%$, which is higher than the fraction derived by \citet{Evans2006}. \newline
These findings combined reveal that the stellar content of the core and the outskirts are clearly different, with the stars in the outskirts significantly younger.  We note this is the opposite of what was found in 30 Doradus, where the core cluster (R136) is significantly younger than the average age of the massive stars in NGC\,2070 and the wider region \citep{Crowther2016, Schneider2018a}.\newline
The two distinct populations seen in/around NGC\,330 could be explained by different star-formation histories, with the older cluster contrasting with younger stars formed as part of the ongoing (more continuous) star formation in the SMC bar. Indeed, given the 35-40 Myr age of the cluster, feedback from when the most massive stars in the cluster exploded as supernovae could have helped instigate star formation in the outskirts.\newline
Alternatively, rejuvenated runaway or walkway stars \citep[see e.g.,][]{Renzo2019} could contribute to the younger population in the outskirts. These stars would be the mass gainers in previous binary interactions (thus appearing rejuvenated) that had been ejected when their previously more massive companions exploded as supernoave. Assuming a low ejection velocity of $5\,\mathrm{km/s}$ a star would travel 50\,pc within 10\,Myr which indicates that the population of the outskirts (i.e. outside the MUSE FoV) by runaway or walkaway stars is possible.\newline
More detailed study is required to understand the formation history of these populations, supported by detailed physical parameters of the stars and combined with dynamical information from their radial velocities and proper motions.

\section{Summary and future work}\label{Sec:conclusion}
This is a first study of the dense core of the SMC cluster NGC~330 using the MUSE-WFM equipped with the new AO. Using the unprecedented spatial resolution of MUSE WFM-AO allows us to spectroscopically study the massive star population in the dense core of the cluster for the first time. We automatically extract and normalize spectra for more than 250 stars. We find that a vast majority of them are B-type stars while there are also two Oe stars and a handful of RSGs and BSGs. Around 30\% of the B stars (and both O stars) show broad spectral lines characteristic of rapid rotation and emission features primarily in H$\alpha$ but also in other spectral lines like He I and O I. They are thus classified as classical Be (Oe) stars. We derive a lower limit on the observed binary fraction of $14\,\pm\,2\%$.\newline
Using archival HST data we are able to position our stars in a color-magnitude diagram and investigate their evolutionary status. Comparing to single-star non-rotating Padova isochrones we estimate a cluster age between 35 and 40 Myr and a total cluster mass of $M_\mathrm{tot,NGC330} = 88^{+17}_{-18} \times 10^3 \,M_{\odot}$. This mass can be used to compare our findings for NGC~330 to clusters of similar masses but different ages, for example RSGC~01, RSGC~02, RSGC~03, or NGC~6611 in the Milky Way, and NGC~1818, NGC~1847 or NGC~2100 in the LMC \citep{Portegies2010}. \newline
We find several stars that are hotter and brighter than the main-sequence turn-off, indicative of blue straggler stars which are thought to be binary-interaction products. Two of these stars, classified as Oe/Be stars are the stars of earliest spectral type in our sample and could have been spun up to high rotational velocities in previous binary interactions.\newline
Comparing our findings to the spectroscopic study of 120 OB stars in the outskirts of NGC~330 by \citet{Evans2006} we find significant differences in the spectral type distribution, the Be star fraction and the binary fraction. We conclude that the stellar content of the core is significantly different from the one in the outskirts and discuss possible explanations.\newline
In subsequent papers of this series we will determine the current multiplicity fraction of all stars in the sample using the 6 epochs we have available. This will also allow us to characterize the variability of the Be stars in the sample. Subsequently, we will combine all 6 epochs to increase the S/N of the data which will allow us to model the stars in order to derive their effective temperatures, surface gravities, rotation rates, and abundances for the brightest stars. This will allow us to further investigate the BiP candidates identified in this paper and characterize the star formation history of NGC~330.

\begin{acknowledgements}
J.B. acknowledges support from the FWO\_Odysseus program under project G0F8H6N.
L. R. P acknowledges support from grant AYA2015-68012-C2-1-P from the Spanish Ministry of Economy and Competitiveness (MINECO).
SdM acknowledges funding by the European Union’s Horizon 2020 research and innovation programme from the European Research Council (ERC) (Grant agreement No. 715063), and by the Netherlands Organisation for Scientific Research (NWO) as part of the Vidi research program BinWaves with project number 639.042.728.
This research has made use of the SIMBAD database, operated at CDS, Strasbourg, France and of NASA's Astrophysics Data System Bibliographic Services.
Parts of the analysis in this project are based on the python code \textsc{photutils}.
We are grateful to the staff of the ESO Paranal Observatory for their technical support.
\end{acknowledgements}

\bibliographystyle{aa}
\bibliography{papers}

\begin{appendix}
\section{Spectral extraction in crowded regions}\label{Sec_appendix:specextract}
\begin{figure*} \centering
\includegraphics[width=0.99\hsize]{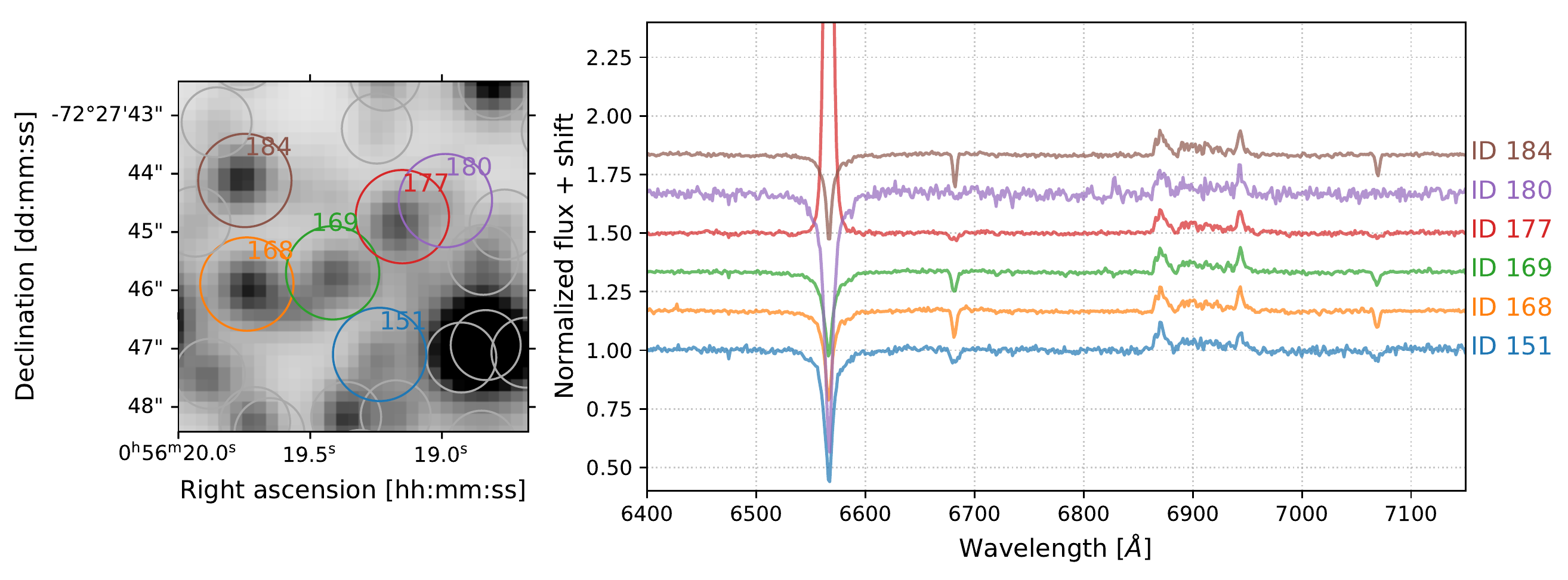}
\caption{Example spectral extraction for star \#169. All stars within 12 pixels distance from the target star (marked with colored circles in the left panel) are taken into account in the extraction of the spectrum. The spectra of all considered stars are shown in the right panel. It can be seen that the spectrum of ID 169 does not show signs of contamination by its surrounding stars.}
\label{fig:spec_extract}
\end{figure*}

Due to the crowding of the observed field, the extraction of spectra is a non-trivial task. We develop a spectral extraction routine based on the \textsc{python} package \textsc{photutils} that is based on PSF extraction. For each star, we take all sources within a distance of 12 pixels (corresponding to 2.4'') into account by simultaneously fitting their PSF. One example is shown in Fig. \ref{fig:spec_extract} where the spectrum of the star \#169 is extracted.

\section{Be star spectral classification}\label{Sec:Bestar_specclass}
\begin{figure*} \centering
\includegraphics[width=0.99\hsize]{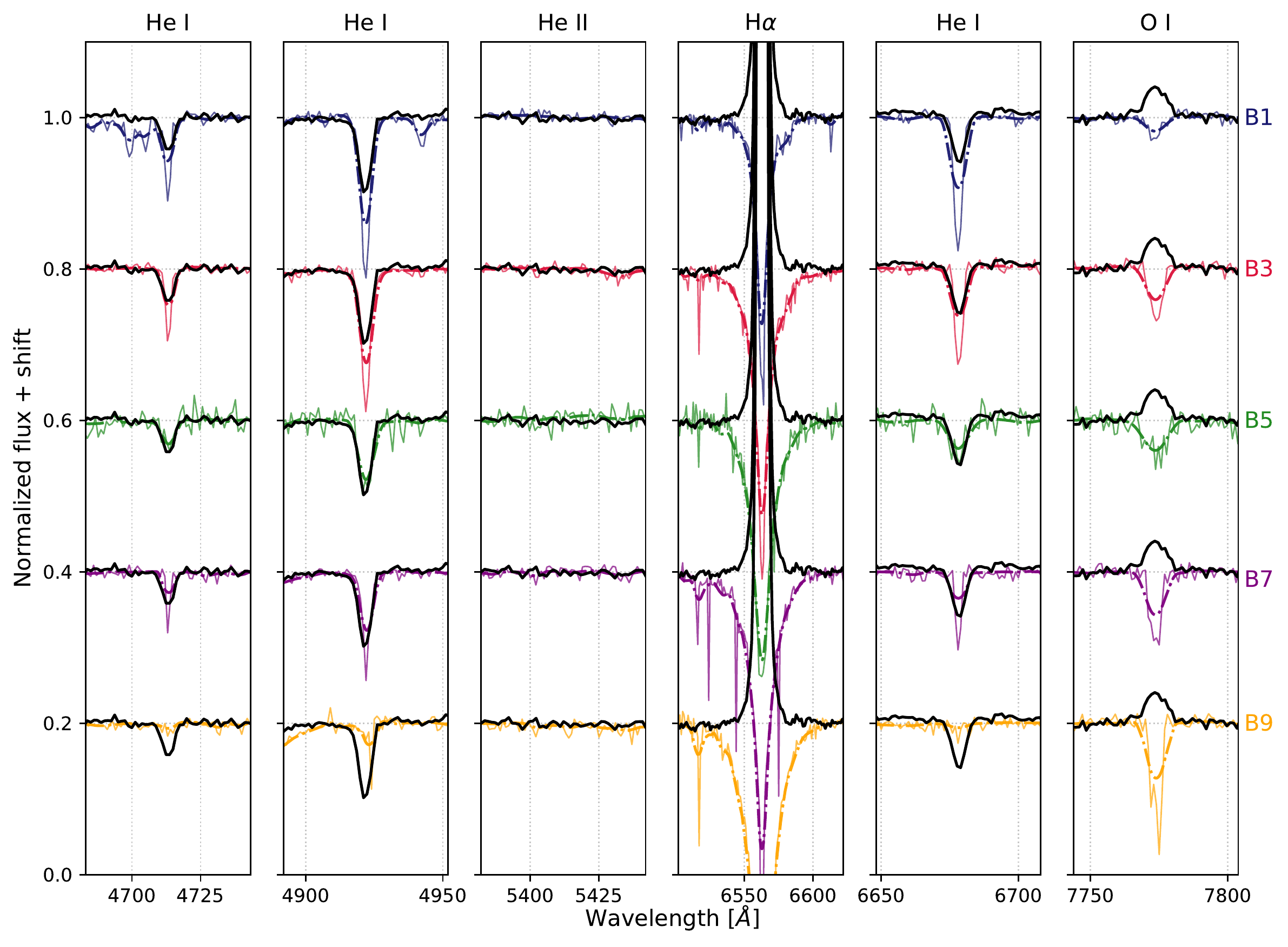}
\caption{Example Be star spectral classification of star \#104. The MUSE spectrum of the star that has to be classified is shown in black. Overplotted and shifted vertically for clarity are a set of standard star spectra (as described in Sect. \ref{Sec:specclass_MS}) of spectral types B1 to B9, degraded to MUSE resolution. While the continuous line shows the observed standard star spectrum the dashed dotted line is rotationally broadened by 200 km/s.}
\label{fig:Be_specclass}
\end{figure*}

As described in Sect. \ref{Sec:specclass_MS}, our automated spectral classification method reveals its limitations for Be stars because of the infilling of spectral lines by emission from the Be star disk. In addition to the automatic classification we thus assign spectral types to the Be stars by comparing their spectra to a set of standard star spectra by eye. We account for the possible rapid rotation of the Be stars by artifically broadening the standard star spectra to a rotational velocity of 200 km/s. Furthermore we rebin them to lower resolution to resemble the MUSE spectra. One example of such a classification is shown in Fig. \ref{fig:Be_specclass}. The star shown here is classified as a B5e star. \newline
For a majority of the Be stars in our sample the two methods agree within the errors. In cases of disagreement we keep the spectral type derived by visual inspection as this method allows us to take more spectral lines into account.

\section{Color-Magnitude Diagram}\label{Sec:appendixCMD}
As mentioned in Sect. \ref{Sec:cmd} none of the isochrones reproduces the lower part of the main sequence, i.e. mainly stars fainter than $F814W = 20\,\mathrm{mag}$. While they reproduce well the upper part of the MS, the isochrones are systematically too faint by $\approx 0.3\, \mathrm{mag}$ around the MS kink. \newline
A shift to brigther magnitudes is equivalent to assuming a lower distance to the cluster (i.e. a distance of $\approx 48\,\mathrm{kpc}$). Given that the distance to the SMC and hence the distance to NGC~330 we refrain from varying the distance.\newline
Varying the extinction towards the SMC does not remove this discrepancy either as the extinction affects the $F336W$ and $F814W$ filters differently and thus changes the shape of the isochrones on the CMD.
\begin{figure} \centering
\includegraphics[width=0.99\hsize]{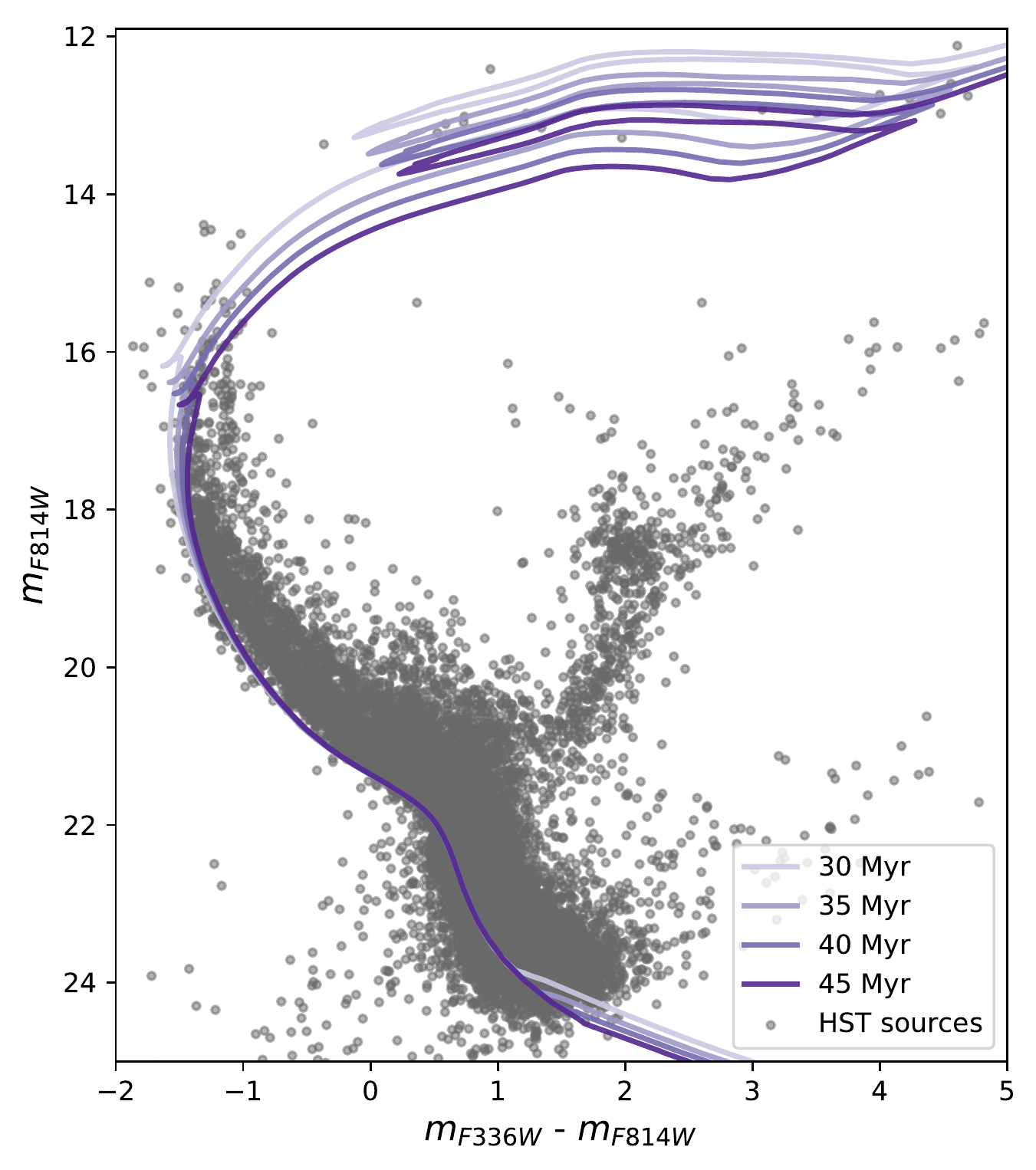}
\caption{Zoomed-out version of Fig.~\ref{fig:CMD} including stars fainter than $F814W = 20\,\mathrm{mag}$. It can be seen that none of the isochrones reproduces the lower part of the MS.}
\label{fig:CMD_zoomout}
\end{figure}

\section{Tables with Equivalent Width measurements}\label{app:table_spectype}
Table \ref{tab:all_params} gives an overview over the coordinates, V-band magnitudes, the measured equivalent width with error for all B type stars, spectral types, and possible comments.

\tiny

\onecolumn
\begin{longtable}{llllrrrlll}
\caption{Compilation of all parameters derived for the sample stars. The first, second, and third columns give the identifier as well as coordinates. The column 'V' gives the apparent V magnitude (derived as described in Sect. \ref{Sec:inputlist}). We then give an overview over the measured equivalent widths, equivalent width errors, and derived spectral types for all B-type stars. For Be stars, we give the spectral type derived with the by-eye method. For the RSGs and BSGs we given the information shown already in Tab. \ref{tab:RSGBSG}. The last column gives comments.}
\label{tab:all_params}\\
\hline
Identifier & RA & DEC & V &  $\mathrm{EW}_{5412}$ & $\mathrm{EW}_{6678}$ & $\mathrm{EW}_{7774}$ & Class & SpType & Comments \\
BSM \# & [J2000]& [J2000]  & [mag] & [$\AA$] & [$\AA$] & [$\AA$] & & & \\ \hline
\endfirsthead
\caption{continued.}\\
\hline
Identifier & RA & DEC & V &  $\mathrm{EW}_{5412}$ & $\mathrm{EW}_{6678}$ & $\mathrm{EW}_{7774}$ & Class & SpType & Comments \\
BSM \# & [J2000]& [J2000]  & [mag] & [$\AA$] & [$\AA$] & [$\AA$] & & &  \\  \hline
\endhead
\hline
\endfoot
001 & 14.053752 & -72.472051 & 18.28 & 0.057 $\pm$ 0.074 & 0.573 $\pm$ 0.070 & -0.138 $\pm$ 0.101 & B & B4 & - \\
002 & 14.071623 & -72.471556 & 18.22 & -0.018 $\pm$ 0.065 & 0.351 $\pm$ 0.068 & 0.008 $\pm$ 0.085 & B & B6 & - \\
003 & 14.093478 & -72.470921 & 17.53 & - & - & - & Be & B5e & - \\
004 & 14.065192 & -72.470558 & 18.26 & 0.049 $\pm$ 0.055 & 0.295 $\pm$ 0.078 & -0.053 $\pm$ 0.084 & B & B6 & - \\
... & & & & & & & & & \\
042 & 14.07937325 & -72.46877809 & 13.50 & - & - & - & RSG & K1 Ib & Arp II-18 \\
... & & & & & & & & & \\
\end{longtable}
\tablefoot{A full version of this table is available electronically. The first few lines are shown as an example.}

\end{appendix}
\end{document}